\documentclass[12pt,preprint]{aastex}
\usepackage{graphicx,color,natbib}

\newcommand{\kms}{km\,s$^{-1}$} 
 
\slugcomment{Published in ApJ, 718 1200}

\shorttitle{Study of New Cold Disks from the c2d Observations}
\shortauthors{Mer\'{\i}n et al.}

\begin{document}


\title{A Spitzer c2d Legacy Survey to Identify and Characterize Disks with Inner Dust Holes}


\author{
Bruno Mer\'{\i}n\altaffilmark{1,3}, 
Joanna M. Brown\altaffilmark{2},
Isa Oliveira\altaffilmark{3},
Gregory J. Herczeg\altaffilmark{2},
Ewine F. van Dishoeck\altaffilmark{2,3},
Sandrine Bottinelli\altaffilmark{4},
Neal J. Evans II\altaffilmark{5},
Lucas Cieza\altaffilmark{6},
Loredana Spezzi\altaffilmark{7},
Juan M. Alcal\'a\altaffilmark{8},
Paul M. Harvey\altaffilmark{5},
Geoffrey A. Blake\altaffilmark{9},
Amelia Bayo\altaffilmark{10},
Vincent G. Geers\altaffilmark{11}
Fred Lahuis\altaffilmark{12,3},
Timo Prusti\altaffilmark{7},
Jean-Charles Augereau\altaffilmark{4},
Johan Olofsson\altaffilmark{4},
Frederick M. Walter\altaffilmark{13},
Kuenley Chiu\altaffilmark{14}}
\altaffiltext{1}{Herschel Science Centre, European Space Astronomy Centre (ESA), 
P.O. Box 78, 28691 Villanueva de la Ca\~nada (Madrid), Spain}
\altaffiltext{2}{Max-Plank Insitut F\"ur Extraterrestrial Physik, Giessenbachstra{\ss}e, 85748 Garching bei M{\"u}nchen, Germany}
\altaffiltext{3}{Leiden Observatory, Leiden University, P.O. Box 9513, 2300 RA Leiden, The Netherlands }
\altaffiltext{4}{Laboratoire d'Astrophysique de Grenoble, Universit\'e Joseph Fourier, CNRS, UMR 5571, Grenoble, France}
\altaffiltext{5}{Department of Astronomy, University of Texas at Austin, 
	    1 University Station C1400 Austin, TX 78712-0259, USA}
\altaffiltext{6}{Institute for Astronomy, University of Hawaii at Manoa, Honolulu, HI 96822. USA}
\altaffiltext{7}{RSSD, European Space Agency (ESTEC), PO Box 299, 2200 AG Noordwijk, The Netherlands }
\altaffiltext{8}{INAF - Osservatorio Astronomico di Capodimonte, Salita Moiariello, 16, 80131 Napoli, Italy}
\altaffiltext{9}{Division of Geological and Planetary Sciences, MS 150-21, 
	    California Institute of Technology, Pasadena, CA 91125, USA}
\altaffiltext{10}{European Southern Observatory, Alonso de Cordova 3107, Vitacura, Santiago, Chile}
\altaffiltext{11}{Astronomy Department, University of Toronto, Canada}
\altaffiltext{12}{SRON Netherlands Institute for Space Research, P.O. Box 800, 9700 AV Groningen, The Netherlands}
\altaffiltext{13}{Department of Physics and Astronomy, Z-3800, Stony Brook University, Stony Brook, NY 11794-3800, USA}
\altaffiltext{14}{Department of Astronomy, MS 249-17, 
	    California Institute of Technology, Pasadena, CA 91125, USA}

\begin{abstract}

  Understanding how disks dissipate is essential to studies of planet
  formation. However, identifying exactly how dust and gas dissipates
  is complicated due to difficulty in finding objects clearly in the
  transition of losing their surrounding material.  We use {\it
    Spitzer} IRS spectra to examine 35 photometrically-selected
  candidate cold disks (disks with large inner dust holes).  The
  infrared spectra are supplemented with optical spectra to determine
  stellar and accretion properties and 1.3mm photometry to measure
  disk masses.  Based on detailed SED modeling, we identify 15 new
  cold disks.  The remaining 20 objects have IRS spectra that are
  consistent with disks without holes, disks that are observed close
  to edge-on, or stars with background emission.  Based on these
  results, we determine reliable criteria for identifying disks with
  inner holes from {\it Spitzer} photometry and examine criteria
  already in the literature.  Applying these criteria to the c2d
  surveyed star-forming regions gives a frequency of such objects of
  at least 4\% and most likely of order 12\% of the YSO population
  identified by {\it Spitzer}.

  We also examine the properties of these new cold disks in
  combination with cold disks from the literature. Hole sizes in this
  sample are generally smaller than for previously discovered disks
  and reflect a distribution in better agreement with exoplanet
  radii. We find correlations between hole size and both disk and
  stellar masses. Silicate features, including crystalline features,
  are present in the overwhelming majority of the sample although 10
  \micron{} feature strength above the continuum declines for holes
  with radii larger than $\sim$7 AU. In contrast, PAHs are only
  detected in 2 out of 15 sources. Only a quarter of the cold disk
  sample shows no signs of accretion, making it unlikely that
  photoevaporation is the dominant hole forming process in most cases.

\end{abstract}





\section{Introduction}

Near- and mid-IR observations of young stars demonstrate
that optically-thick circumstellar disks disappear around
approximately half of low-mass young stars in 1-3 Myr and are nearly
entirely absent around members of 10 Myr old associations
(e.g. \citealt{Haisch2001}; \citealt{Gutermuth2004};
\citealt{Sicilia-Aguilar2005}, \citealt{Low2005};
\citealt{Currie2009}).  Accretion ceases on approximately the same
timescale (e.g. \citealt{Calvet2005}). The disappearance of gas and
dust - planetary building material - places stringent limits on the
timescales of giant planet formation
\citep{Pollack1996,Kenyon2009}. However, identifying exactly how dust
and gas dissipate is complicated due to the difficulty in identifying
objects clearly in the transition of losing their surrounding
material.

One of the most common methods, dating back to IRAS
(e.g. \citealt{Strom1989,Skrutskie1990}), of identifying candidate
transitional systems between the classical and debris disk
evolutionary stages is through mid-IR spectral energy distributions
(SEDs). Dust growth, sedimentation and removal is expected to
result in a deficit of infrared flux as material is accreted or
dispersed. This deficit has become the defining characteristic of
transitional disks.  For some disks, flux deficits are seen at all
wavelengths, suggesting a gradual dissipation of mass for all disk
radii.  In other cases, a flux deficit is only seen at short
wavelengths, indicating that the outer disk remains massive and
optically thick while the inner disk lacks small dust grains.  The presence of both
types of transitional disks suggests that the transition happens via
different paths from the gas-rich, optically-thick stage to gas-poor,
optically-thin disks \citep{Cieza2007,Currie2009}. The fraction of
stars with transitional disks is thought to be 5--25\%
\citep{Lada2006,Hernandez2007,Sicilia-Aguilar2008,Dahm2009,Currie2009,Kim2009}.
The small numbers indicate that the evolutionary path through a
transitional disk is either uncommon or rapid. However, a consensus in
nomenclature is still lacking: even within a single cluster, the
calculated fraction of transitional disks can range from 10-50 \%,
depending on definitions (\citealt{Ercolano2009};
\citealt{Sicilia-Aguilar2008}).

Transition disks with a short wavelength deficit are particularly
interesting because they are potential tracers of planet
formation. This deficit arises from the absence of hot small dust
grains close to the star resulting in flux coming solely from the
stellar photosphere, rather than disk surface emission.  This deficit
may be the result of grain growth and sedimentation but also of
complex interactions between a nascent protoplanet and the surrounding
disk.  The flux deficit from such inner holes results in a depressed
SED at wavelengths less than $\sim$15 \micron, while retaining typical
fluxes at longer wavelengths from the outer disk.  The short
wavelength flux deficit arises from the absence of hot small dust
grains close to the star resulting in flux coming solely from the
stellar photosphere, rather than disk surface emission. We define such
disks as ``cold disks'' to differentiate from other types of
transitional systems (e.g. anemic or homologously depleted, defined as
disks with an overall lower mass at all radii) and to emphasize that
this is an observational characteristic not necessarily tied to
evolution.

SEDs, relying heavily on mid-IR spectroscopy, are the tool
currently most widely used to infer the presence of holes and gaps
(e.g. \citealt{Calvet2002, Forrest2004, Brown2007}, hereafter
B07). Additional submillimeter observations of the most massive disks
have directly imaged large holes \citep{Pietu2006, Hughes2007,
  Hughes2009, Brown2008, Brown2009, Andrews2009}. The hole sizes match
reasonably well with estimates from SED modeling, suggesting that
current interpretation and modeling of the SEDs is correct. Within the
category of cold disks, significant differences are seen in hole sizes
(from 1--50 AU), the presence or absence of gas and dust within the
hole, and the presence or absence of accretion
\citep{Najita2007,Espaillat2007b, Pontoppidan2008,Salyk2009, Kim2009}.
Detailed comparisons of these properties to the predictions of each of
the scenarios listed above can help to discriminate which processes
control the observed disk evolution.  However, statistical studies
\citep[e.g.][]{Najita2007} are limited by small sample sizes. {\it
  Spitzer} mapping efforts should include a significant number of
unidentified cold disks and other types of transitional
disks. However, accurately identifying these cold disks out of the
large numbers of stars present in the maps is difficult with just
broad band photometry. Reliable criteria for identifying disks with
inner holes from {\it Spitzer} photometry, which lack any information
on fluxes in the crucial region between 8 and 24 \micron, are needed
to generate a large sample for statistical purposes.

This paper presents a new large sample of cold disks identified from
`Cores to Disks' data \citep{Evans2009} and analyzes simultaneously follow-up
observations of the sample with the IRS and MIPS instruments together
with optical spectroscopy and millimeter continuum observations.  It
is organized as follows: \S~\ref{obs} describes the target selection,
observations and data reduction of the different data sets, including
the {\it Spitzer}, optical spectroscopy and millimeter
observations. The results are presented in \S~\ref{stellarparameters}
and \S~\ref{diskresults}, describing, in order, the stellar parameters
in \S~\ref{stellarparameters}, disk masses in \S~\ref{diskmass}, the disk
parameters in \S~\ref{diskparameters} and the dust mineralogy in
\S~\ref{mineralogy} for the whole sample.  Notes on individual sources
are given in \S~\ref{individualnotes}. The combined analysis is
presented at \S~\ref{discussion}, with first the presentation of a new selection
criteria for transitional disks in \S~\ref{selection}, then a
description of the observational properties of our cold disks in
\S~\ref{colddiskproperties}, and finally a discussion on the possible
origins of the inner holes in \S~\ref{holeorigins}.  The conclusions
of this work are given in \S~\ref{conclusions}.

\section{Observations and data reduction}
\label{obs}
\subsection{Target selection}
\label{targetsel}

The c2d {\it Spitzer} Legacy program completed a full IRAC (3.6 to 8
$\mu$m) and MIPS (24 to 160 $\mu$m) survey of five nearby star-forming
regions (Perseus, Chameleon II, Lupus, Ophiuchus, and Serpens)
\citep{Evans2003,Jorgensen2006,Alcala2008,Merin2008,Harvey2007b}.
From these maps, 1024 Young Stellar Object (YSO) candidates were
identified using the selection techniques, including removing
extragalactic contaminants, described in \citet{Harvey2007b} (see
\citealt{Evans2009} for a complete description of this sample and a
general study of cloud to cloud differences). This sample should
include a significant number of cold disks but they must first be
separated from the bulk of the population.
   
Cold disks were selected for IRS follow-up using the following
method. Spectral types were taken from the literature where possible
or otherwise a K7 photosphere was used to get an initial estimate of
the SED. Sources with photospheric fluxes in at least IRAC 1 (3.6
$\mu$m) and IRAC 2 (4.5 $\mu$m) and 24 $\mu$m excesses equal or
greater than IRAC 4 (8.0 $\mu$m) fluxes in $\lambda$F$_{\lambda}$
space were selected. Sources with rising or flat SEDs from 8 $\mu$m
to 24 $\mu$m were preferentially included. A flux lower limit of 15
mJy at 8 $\mu$m was imposed to further reduce extragalactic contamination and
keep integration times reasonable. The initially selected sample was
then cut slightly to reject likely contaminants and to avoid
overlap with other IRS programs. A total of 33 sources were selected
in this fashion. To further expand our sample, two additional
candidate cold disks within the c2d clouds were included from a survey
of Weak Line T Tauri stars (WTTs, \citealt{Padgett2006,Cieza2007})
based on their IRS spectra suggestive of inner holes. In total, 35
objects were included in our sample and followed-up with IRS and MIPS
(\S~\ref{spitzerobs}), optical spectroscopy (\S~\ref{opticalobs}), and
millimeter continuum observations (\S~\ref{mmobs}).

\label{observations}

\subsection{Spitzer IRS and MIPS observations}
\label{spitzerobs}

Spectra for the 35 objects in Table \ref{tab:sample} were obtained
using the Infrared Spectrograph (IRS) aboard the {\it Spitzer} Space
Telescope under a variety of programs. Most were observed as part of
program 30843 (P.I.: B. Mer\'{\i}n), which also included deep 70 $\mu$m MIPS
staring observations of the same sample. Objects 17 and
19 were observed with IRS as part of the c2d second look observations
of WTTs from \cite{Padgett2006} and object 25 was observed during the
IRS Guaranteed Time Observation (GTO) although it was selected as a
candidate from the c2d photometry. All the candidate cold disks in Serpens were
observed as part of a complete flux-limited IRS survey in Serpens, program 30223
(P.I.: K. Pontoppidan) and will also be discussed along with the rest of
the Serpens sample in \citet{Oliveira2010}.  Table \ref{tab:sample} gives
the AORs and observing dates for both the IRS and MIPS-70 observations
of the sample.

\begin{figure*}[!ht]
    \includegraphics{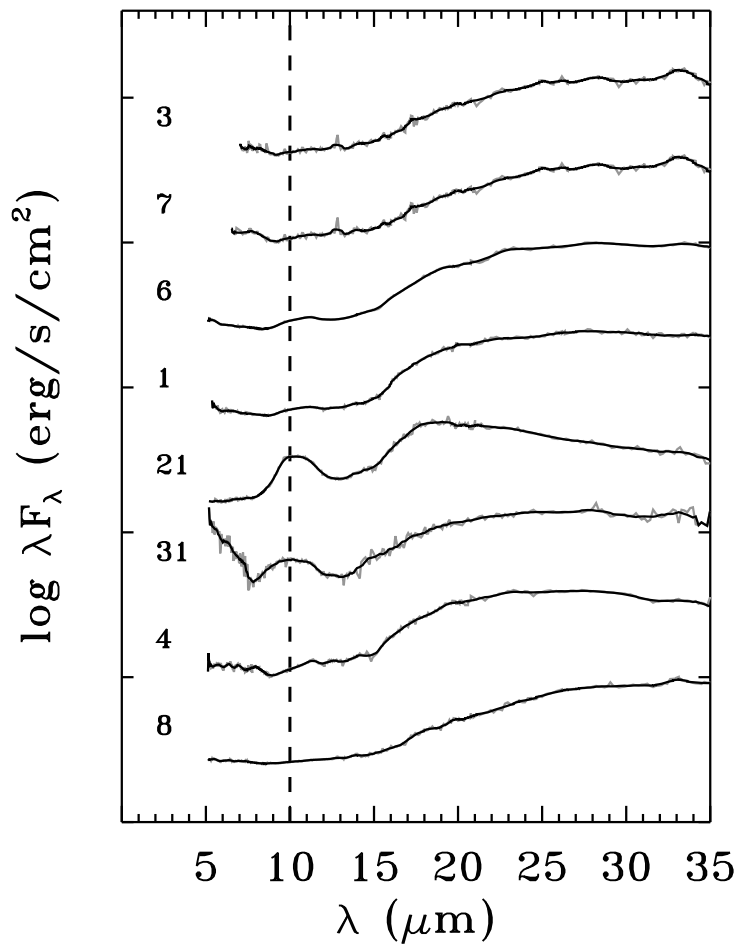}
\caption{IRS spectra of the of the close to edge-on systems ordered
from bottom to top with growing effective temperature. The thick and grey
lines are the binned and original spectra, respectively and the numbers
give their identifications in Table \ref{tab:sample}. \label{IRS_edgeon}}
\end{figure*}

\begin{figure*}[!ht]
    \includegraphics[width=\textwidth]{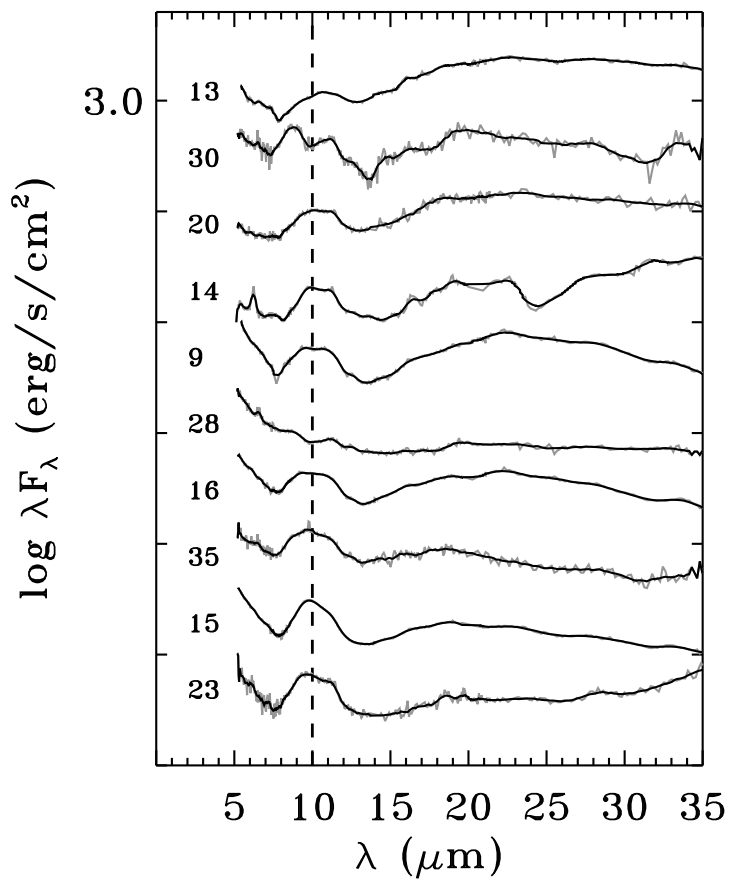}
\caption{IRS spectra of the disks in the sample for which inner holes were not required to obtain a good SED fit organized 
from bottom to top with increasing SED slope $\alpha$. The thick and grey
lines are the binned and original spectra, respectively and the numbers
give their identifications in Table \ref{tab:sample}. \label{IRS_noholes}}
\end{figure*}

\begin{figure*}[!ht]
    \includegraphics[width=\textwidth]{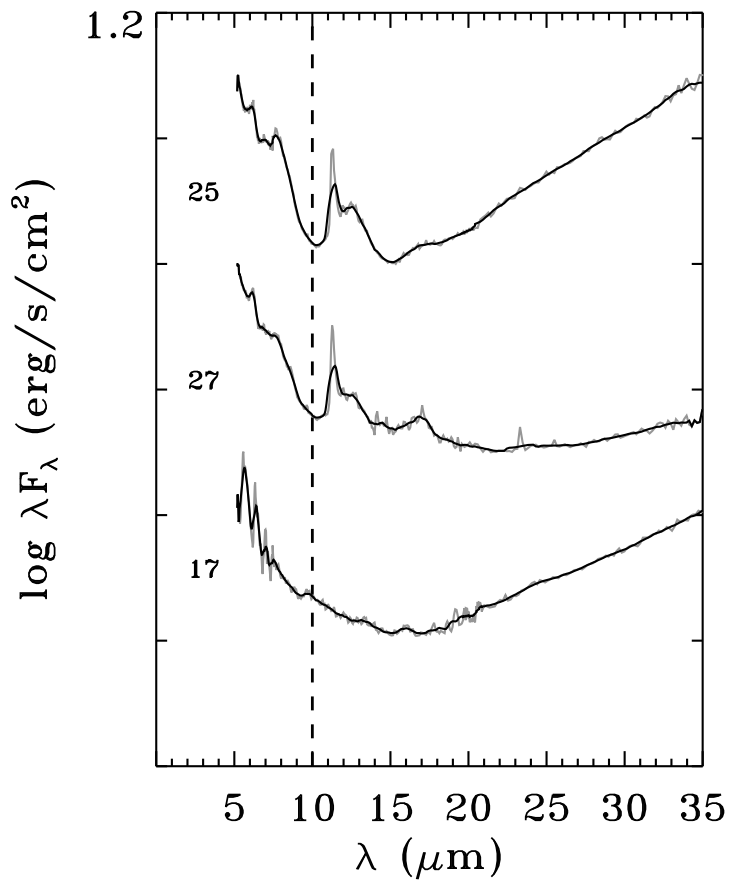}
\caption{IRS spectra of the disks with potentially extended mid-IR emission from surrounding clouds ordered from bottom to top
  with growing stellar mass. Object 17 (Sz 84) is shown here as comparison due to its similar
  IRS spectrum, although it is classified as a cold disk. The thick and grey
lines are the binned and original spectra, respectively and the numbers
give their identifications in Table \ref{tab:sample}.  \label{IRS_ext}}
\end{figure*}

\begin{figure*}[!ht]
    \includegraphics[width=\textwidth]{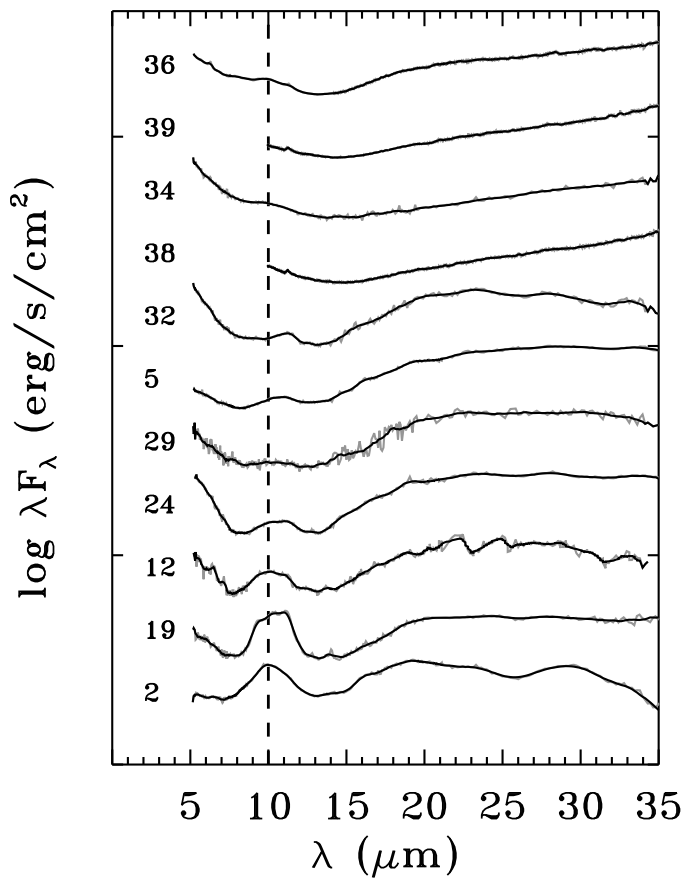}
\caption{IRS spectra of the accreting cold disks in the sample ordered from bottom to top
  with growing inner hole radii. The thick and grey
lines are the binned and original spectra, respectively and the numbers
give their identifications in Table \ref{tab:sample}.  \label{IRS_CD_acc}}
\end{figure*}

\begin{figure*}[!ht]
    \includegraphics[width=\textwidth]{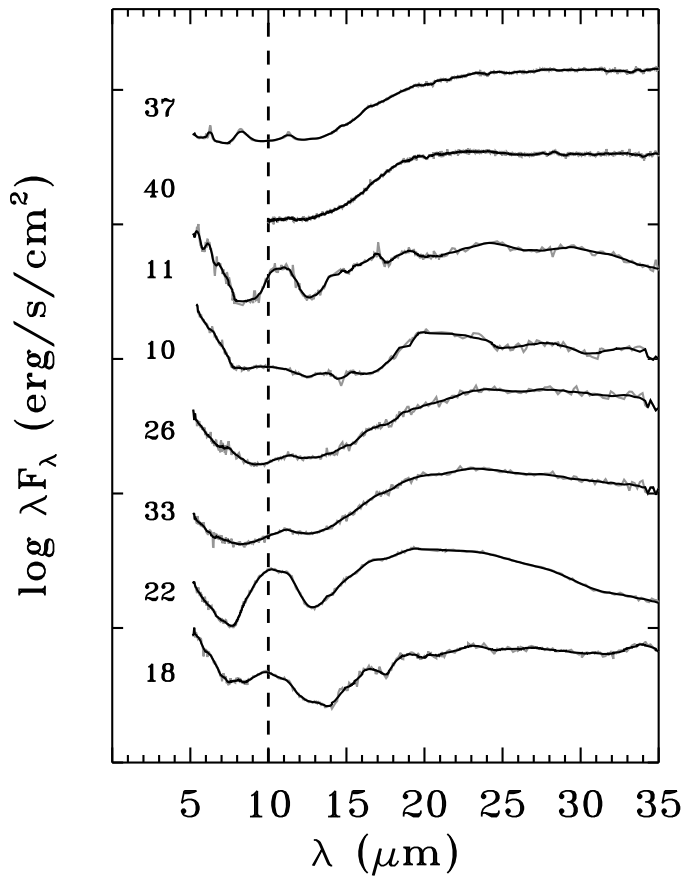}
\caption{IRS spectra of the non-accreting or non-classified cold disks ordered from bottom to top
  with growing inner hole radii. The thick and grey
lines are the binned and original spectra, respectively and the numbers
give their identifications in Table \ref{tab:sample}.  \label{IRS_CD_noacc}}
\end{figure*}

All the IRS spectra were obtained with a combination of Short-Low (SL)
and Long-Low (LL) modules, which provide a resolving power of $R =
\Delta\lambda / \lambda \sim 100$ and a combined wavelength coverage
from 5.3 to 35.0 $\mu$m. Integration times were estimated from the
IRAC flux at 8 $\mu$m and MIPS flux at 24 $\mu$m. The observations
typically ranged from 5 to 20 minutes, resulting in signal to noise
values of 30 to 150, depending on the background level and the
brightness of the objects. Given the rising SED of the targets at long
wavelengths, longer times were usually needed in the SL than LL
observations. The observations were scheduled in cluster mode,
grouping nearby sources with similar integration times to increase the
observing efficiency. Data reduction started from the Basic Calibrated
Data (BCD) images, pipeline version S12.4.0. The processing includes
bad-pixel correction, extraction, defringing, and order matching using
the c2d analysis pipeline (\citealt{Kessler-Silacci2006}, see also the
c2d Spectroscopic Explanatory Supplement, from the SSC
webpage\footnote{\tt http://ssc.spitzer.caltech.edu/legacy/c2dhistory.html}). The
final spectra can be found in Figures \ref{IRS_edgeon} to
\ref{IRS_CD_noacc}.

The same clustering strategy was used for the MIPS observations at 70
$\mu$m. The integration times were set to 300 seconds, with achieved sensitivities 
given in Table \ref{tab:phottable2}. The small maps were
reduced following the procedure described in \cite{Cieza2008}
and using the SSC median-filtered BCDs from the SSC pipeline version
S16.1.0. Details of the method can be found in the aforementioned
paper, but in short, the photometry is determined with an aperture of
16$''$ radius and a sky annulus with inner and outer radii of 48$''$
and 80$''$, respectively. Visual inspection and correction factors
suggested by the SSC were used in all cases. The resulting fluxes are
given in Table \ref{tab:phottable2}, together with the other IRAC and
MIPS fluxes from c2d.

\subsection{Optical Spectroscopy}
\label{opticalobs}

Optical spectra were obtained for 21 objects in our sample using the
Wide Field Fibre Optical Spectrograph (WYFFOS, \citealt{Bingham1994})
on the 4.2m William Herschel Telescope (WHT), the Intermediate
dispersion Spectrograph and Imaging System (ISIS) at the WHT, the
Intermediate Dispersion Spectrograph (IDS) at the Isaac Newton
Telescope (INT, see Carter et al. 1994\footnote{\tt http://www.ing.iac.es/Astronomy/observing/manuals/man\_wht.html} 
for a description of last three instruments), the Calar Alto Faint Object Spectrograph 
(CAFOS) at the 2.2m Calar Alto Telescope, the Double Spectrograph (DBSP, Oke \& Gunn
1982) at the 5m Hale Telescope at Palomar Observatory, and with the
R-C Spectrograph on the 1.5m CTIO telescope.  These observations
typically cover $\sim 5500-8500$ \AA\ with a resolving power of $\sim
1000-5000$.  Table \ref{t_obs} lists the instrument setups and observation log.

Data were reduced and spectra were extracted using standard methods in
IDL and IRAF.  No telluric correction or flux calibration was
performed.  For WYFFOS, a fiber-fed multi-object spectrograph, sky subtraction 
was obtained from fibers that were placed on blank sky regions (see 
\citealt{Oliveira2009} for details).

\subsection{IRAM-30 millimeter observations}
\label{mmobs}

A subsample of 17 of the northern targets was observed with the
IRAM-30m telescope in Pico Veleta. The observations were carried out
with MAMBO-2 \citep{Kreysa2002} mounted on the IRAM-30m Telescope
during the 2007 winter and summer pool sessions. MAMBO-2 is a
117-bolometer array with a half-power spectral bandwidth of 80 GHz
centered on $\sim$ 250 GHz (1.2 mm), yielding a beam size of 11$''$.
The data were analysed with the MOPSI software \citep{Zylka1998}. The
flux calibration was performed by observing either Mars or Uranus to
determine the flux conversion factor. For each channel the sky noise
was subtracted by computing the weighted mean of the signals from the
surrounding six channels.

All Perseus sources from our sample (see Table \ref{tab:sample}) were
observed, except numbers 3, 6, 8 and 11. We also observed sources 20,
21 and 22 in Ophiuchus and sources 33, 34 and 35 in Serpens. The
ON/OFF observing mode was used with a throw of 35$''$ and each source
was observed until a $>3\sigma$ detection was obtained, or until an
rms of 0.7 mJy (whenever possible) was reached. The integration times
ranged from 10 to 40 minutes.  In general, the sky noise of the array
was low ($<80$ mJy), and was never larger than $\sim$
100 mJy. Ten sources were detected with fluxes between 1 and 10
mJy, while the other 7 have upper limits between 1 and 7 mJy. These
fluxes can be found in column 8 of Table \ref{tab:phottable2}. 

\section{Stellar Properties}
\label{stellarparameters}
Stellar properties must be determined to understand and interpret data
on the surrounding disks.  Optical spectra are used here to measure
spectral types and to determine whether accretion is ongoing.  We
obtained optical spectra for 21 of the 35 objects in our sample. Of
the other 14 stars, 11 have published spectral types and accretion
properties.  For the remaining three sources, spectral types are
estimated from an SED analysis of broadband photometry and accretion
properties are left uncertain.  The SED fits are also used to measure
the extinction to each source.  Results are presented in Table
\ref{tab:stellarparameters}.

For the 21 objects for which we obtained optical spectra, spectral
types of G and K stars were assigned by finding a best match to the
depth of photospheric absorption features in template spectra obtained
from the EXPORT spectral library \citep{Mora2001}.  Spectral types of
M-dwarfs were assigned by finding the best match to the TiO absorption
bands from template spectra \citep{Montes1997}.  The uncertainty in
SpT is typically $\sim 3$ subclasses for G and K stars and $1$
subclass for M stars.  Object 6 is classified as a continuum object
because no spectral features were detected in the low-resolution
spectrum. Any possible optical veiling was not considered for the
other objects in the sample.  A more detailed description of our
spectral typing and the stellar properties of all the Serpens objects
are presented by \cite{Oliveira2009}. Literature values for SpT are
adopted for several objects that we did not observe and for several
objects where the literature SpT is more reliable than the SpT from
our spectra.  Spectral types are converted to $T_{eff}$ using the
scales in \citet{Kenyon1995} for objects
earlier than M0  and \citet{Luhman2003} for those later or equal to M0, to account for the lower surface gravity atmosphers of mid M stars.

The extinction and luminosity for each object were calculated using
SED fits to broadband optical and near-IR photometry (see Table \ref{tab:phottable1} 
for photometry and \S~\ref{diskparameters} for a discussion on the assumed extinction
law).  Assumed distances are listed at the bottom of Table
\ref{tab:stellarparameters}. NEXTGEN models of stellar photospheres
\citep{Hauschildt1999} are used as templates for the broadband
emission.  The 2MASS and any optical photometry are the primary
constraints on extinction and stellar luminosity. For the 3 objects
that lack optical spectra (identified with a ``2'' in Table
\ref{tab:stellarparameters}), the SED fits are also used to estimate
SpT.  However, these SpT estimates have large uncertainties because
they are based on JHK colors, which are not very sensitive to spectral
types from mid-K through M (\citealt{Leggett1992};
\citealt{Kenyon1995}).

Stellar ages and masses are estimated by comparing the stellar
luminosity and effective temperature to pre-main sequence tracks of
\citet{Baraffe1998} for low-mass stars ($<1.4$ $M_\odot$, mixing length = 1) and
\citet{Siess2000} for higher-mass stars.  Since uncertainties in stellar
age are large they are not tabulated here and we focus the analysis on
the better constrained stellar masses.

\subsection{Edge-on Disks}

In the process of determining the stellar properties, it became clear
that a fraction of the sample sources were actually edge-on disks. Inferring
disk structure from an SED analysis is difficult when the disk is
viewed edge-on and occults the star.  In these cases, optical and
near-IR light from the star and inner disk are only seen in reflected
emission \citep[e.g.][]{Padgett1999}.  The strength of such emission
depends on the precise viewing angle and the disk flaring, but is
always much fainter than it would be if the disk were viewed without the 
large absorption column from the edge-on disk.  Emission from outflows, 
seen prominently in optical forbidden lines, are not occulted by the disk 
and can therefore have large equivalent widths \citep{White2004}.

Based on the existing data on our sources, we establish four criteria
to determine whether a disk is viewed edge-on: (1) a low ratio of
photospheric luminosity to main-sequence luminosity, assessed at
$10^{8}$ Myr from Baraffe et al. (1998); (2) large equivalent widths
in optical forbidden lines; (3) strong silicate and/or ice absorption
in the IRS spectra; and (4) high infrared luminosity compared to
stellar luminosity.  Any one of these criteria can be degenerate
between edge-on disks and other explanations, due to potential errors
in distance and extinctions and the absence of accretion/outflow
activity.  However, the presence of at least two of these four
criteria should establish in most cases that the disk is
edge-on. Eight of the sources were selected as edge-on disks and the
reasons are marked in Table \ref{tab:edgeon}. No further analysis was
done on these disks as the stellar properties are too uncertain.

\subsection{Accretion Properties}

The presence or absence of accretion can be assessed from the strength
and shape of emission in several optical lines, most prominently
H$\alpha$.  Though some H$\alpha$ emission is produced in
chromospheres of late-type stars, accretion typically generates
H$\alpha$ emission with much larger equivalent widths and broader
spectral profiles (see \citealt{Hartmann2001} for a comprehensive
description of accretion phenomenology and diagnostics).  A 10 \AA\
cutoff in equivalent width has historically been used to identify
accretion, though this rough cutoff depends on spectral type
(\citealt{White2003}; \citealt{Fang2009}).  More recently, the
H$\alpha$ 10\% width (defined as the full width of the H$\alpha$ line
at 10\% of the peak flux) has been used to discriminate between
accretors and non-accretors (\citealt{White2003};
\citealt{Muzerolle2003}; \citealt{Jayawardhana2006}).
Accretion/outflow activity can also produce asymmetries and absorption
components within the line profile (e.g., \citealt{Reipurth1996},
\citealt{Muzerolle2003}, \citealt{Kurosawa2006}) and variable line
fluxes and shapes (e.g., \citealt{Nguyen2009}).  For any such
criteria, a small accretion rate relative to the stellar luminosity is
not detectable.

To classify the accretion properties of stars in our optical sample, a
Gaussian profile is fit to the H$\alpha$ line.  We then measure the
equivalent width and, when possible, the H$\alpha$ 10\% width
from the Gaussian profile, after accounting for the resolution of each
observation (see a more complete discussion of this method in
\citealt{Oliveira2009}).  Uncertainties in H$\alpha$ 10\% widths range
from 50--150 \kms.  The H$\alpha$ equivalent width and 10\% width are
then used together with the spectral type to assess whether accretion
is ongoing, adhering to the loose guidelines suggested by
\citet{White2003} and \citet{Fang2009}.  Typically, a star is
classified an accretor if the H$\alpha$ 10\% width is $>300$ \kms\
and the H$\alpha$ equivalent width is $>3$ \AA\ for an early K star,
$>10$ \AA\ for late K star and $>20$ \AA\ for an M star.

We classify 21 stars in our sample as accretors and 8 stars as
non-accretors, with 6 cases lacking information.  In many cases an
object is clearly accreting.  As explained below, some classifications
are ambiguous both because some accretors and non-accretors have
similar H$\alpha$ 10\% and equivalent widths and because the
uncertainties in H$\alpha$ 10\% width are large.  Objects 23 and 24
have weak H$\alpha$ equivalent widths but broad H$\alpha$ line shapes.
Such observations could be explained if accretion is present with
strong absorption, either from outflows or the accretion flow itself,
that suppresses the H$\alpha$ emission (e.g. \citealt{Reipurth1996}),
if the star is an unresolved binary with only one accreting component,
or if the stars are chromospherically active with large $v$ sin $i$
\citep{Jayawardhana2006}.  The highest-resolution spectrum of object
24 shows absorption in H$\alpha$, while two of the three show
[\ion{O}{1}] in emission.  This star is therefore classified as an
accretor.  On the other hand, object 23 shows no other evidence of
accretion and is classified as a non-accretor.  Object 18 is left
unclassified because the H$\alpha$ equivalent width is ambiguous
between accretion and non-accretion.  The H$\alpha$ emission from DoAr
21 (object 25) has varied on long timescales, but with no definitive
evidence for accretion \citep{Jensen2009}.  Temporal variations in
H$\alpha$ also introduce uncertainty in classifying the presence or
absence of accretion. A mass-dependent bias likely exists in this
classification scheme because the stronger photospheric continuum
emission and spectral type degeneracy for stars with early spectral
types can mask H$\alpha$ emission produced by weak-to-moderate
accretion onto the star. Based on these uncertainties, we suspect a
small number of objects in our sample (estimated at $<$10\%) might be
misclassified as accretors or non-accretors.

To quantify accretion, we convert the H$\alpha$ 10\% widths to
accretion rate using the relationship derived by \citet{Natta2004}.
Methodological differences in measuring the H$\alpha$ 10\% width
between this work and \citet{Natta2004} likely leads to an
order-of-magnitude uncertainty in the listed accretion rates.  For
late-K and early M-dwarfs, which comprise the bulk of our sample, the
H$\alpha$ 10\% width is not sensitive to accretion rates
$<10^{-10}$ $M_\odot$ yr$^{-1}$.  For earlier spectral types, the
sensitivity worsens to larger accretion rates. Table \ref{tab:stellarparameters}
gives the H$\alpha$ equivalent widths, 10 \% widths and mass accretion
rates derived from this work and Table \ref{tab:sampleclass} gives the
final classification of each source.

\section{Results: Disk and Disk Hole Parameters}
\label{diskresults}

\subsection{Disk mass}
\label{diskmass}

Millimeter continuum fluxes can be used to calculate disk masses
assuming that the millimeter emission is optically thin and a gas to dust 
mass ratio of 100. The disk masses can then be used in turn to constrain SED models.  The disk
masses, $M_{\rm Disk}$, are calculated following \citet{Beckwith1990}
and with standard assumptions and parameters. In the Rayleigh-Jeans
limit,

\begin{equation}
 M_{\rm Disk}={F_\nu}\frac{d^2}{2k<T>}\frac{c^2}{\nu^2\kappa_\nu },
\end{equation} 

where ${F_\nu}$ is the flux at frequency $\nu$, here 230 GHz, $d$ is
the distance to the star (see Table \ref{tab:stellarparameters}), $k$ is the Boltzmann constant,
and $\kappa_\nu$ is the opacity.  It is assumed that the emission
arises primarily from an approximately isothermal region with
temperature, $<$$T$$>$. Average disk temperatures in the outer disk range from 10
to 50 K and here we have used 30 K. One of the major uncertainties is
the opacity, $\kappa_\nu$. We follow \citet{Beckwith1990} and
adopt $\kappa_0$ = 0.02 cm$^2$g$^{-1}$ anchored at $\nu_0$ = 230 GHz,
the frequency of our observations. 

Disk masses are listed in Table \ref{tab:diskparams} for the ten
detected sources and the seven with upper limits.  The majority (80\%)
of the disks in our sample have disk masses of less than 2
 $\times$ 10$^{-3}$ M$_\odot$ and the sample has an average $M_{\rm Disk}$ of
4.5$^{+5.5}_{-1.4} \times 10^{-3}$ M$_\odot$.  These are relatively low
mass disks compared to many millimeter studies, such as the large
sample of T Tauri disks in \citet{Andrews2007} which have an average
mass of 10$^{-2}$ M$_\odot$ using the same assumptions.

\subsection{Disk hole parameters from SED modeling}
\label{diskparameters}

\begin{figure}[h!]
\epsscale{0.6}
\plotone{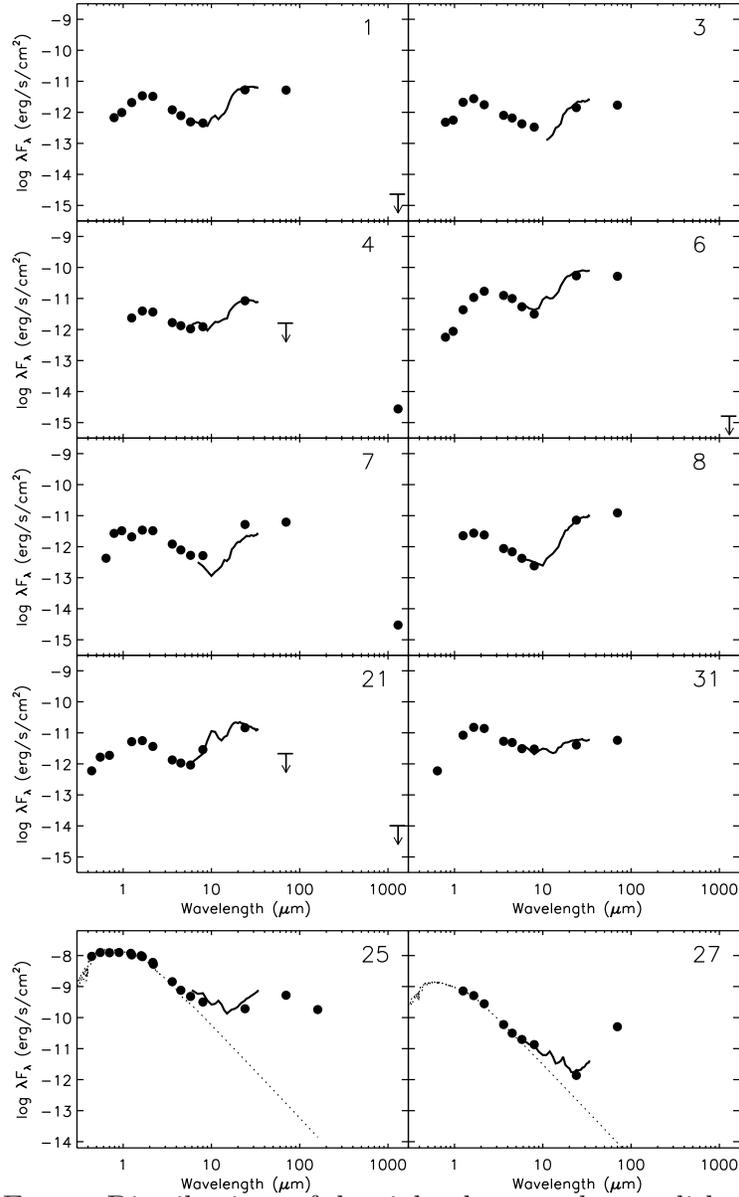}
\caption{Spectral Energy Distributions of the eight close to edge-on disks (top) in the
sample and the two extended sources (bottom). Points are observed fluxes from IRAC, MIPS and IRAM. The
thick solid line is the IRS spectrum and the numbers
give their identifications in Table \ref{tab:sample}. \label{seds_figure_edgeon}}
\end{figure}

\begin{figure}[h!]
    \includegraphics[width=0.8\textwidth]{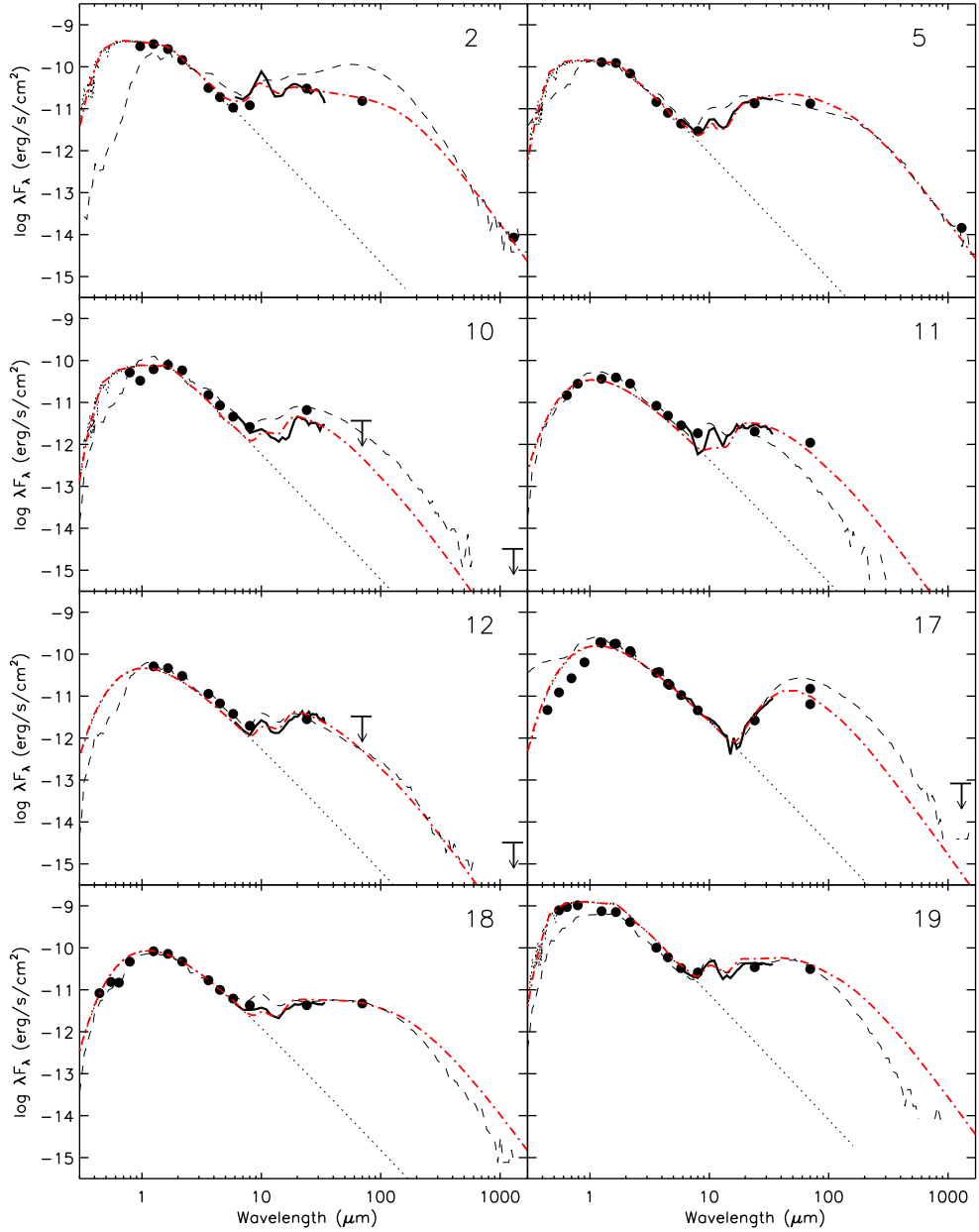}
\caption{Spectral Energy Distributions of the disks with holes in the
sample. Points are dereddened fluxes from IRAC, MIPS and IRAM. The
thick solid line is the IRS spectrum, the dotted line is the stellar photospheric model, the red dash-dot line is the best-fit RADMC disk model, and the thin black dashed line is the Robitaille model fit. The numbers
give their identifications in Table \ref{tab:sample}. \label{seds_figure_holes}}
\end{figure}

\begin{figure}[h!]
    \includegraphics[width=0.8\textwidth]{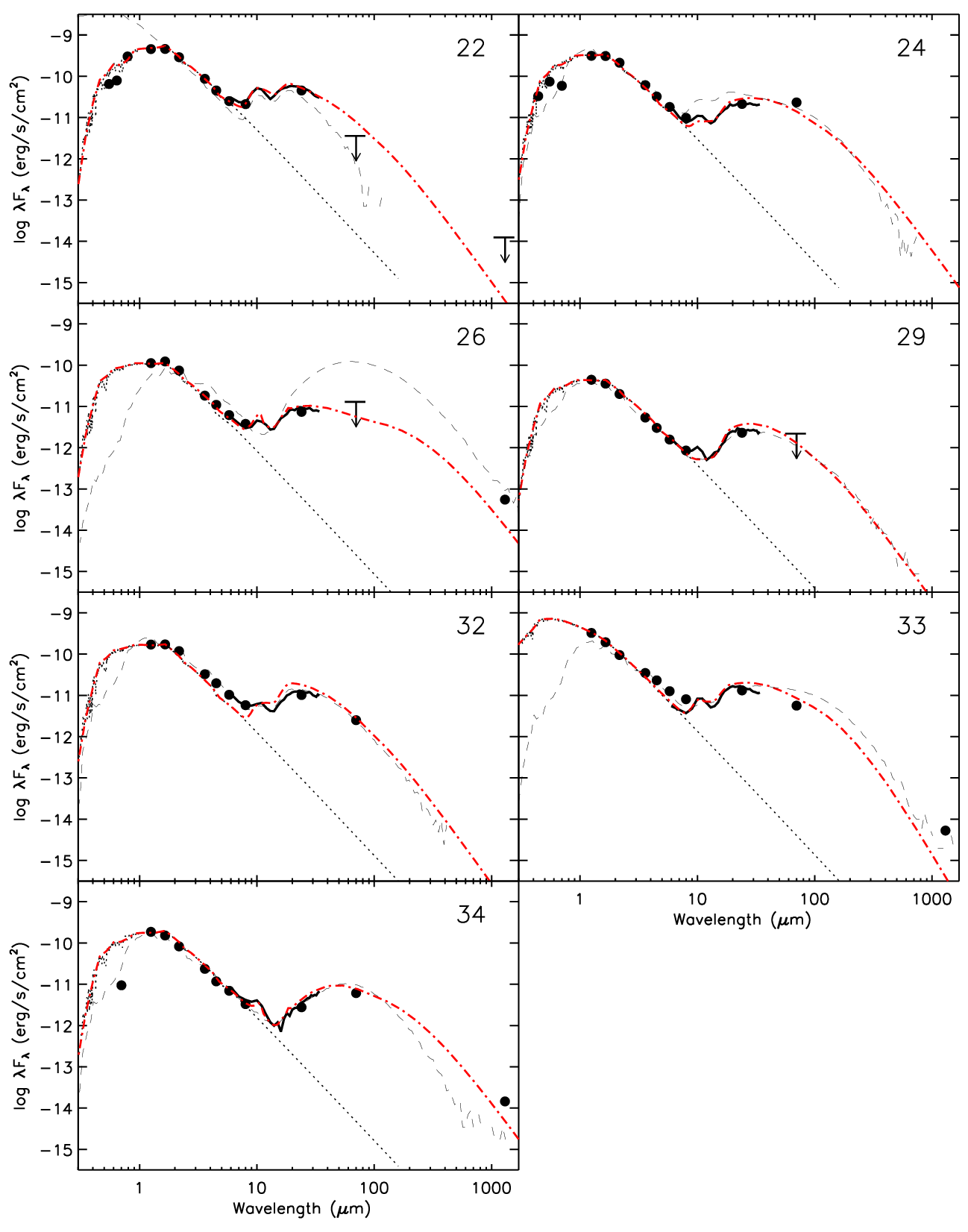}
\caption{Figure 7 continued.\label{seds_figure_holes2}}
\end{figure}

\begin{figure}[h!]
    \includegraphics[width=0.7\textwidth]{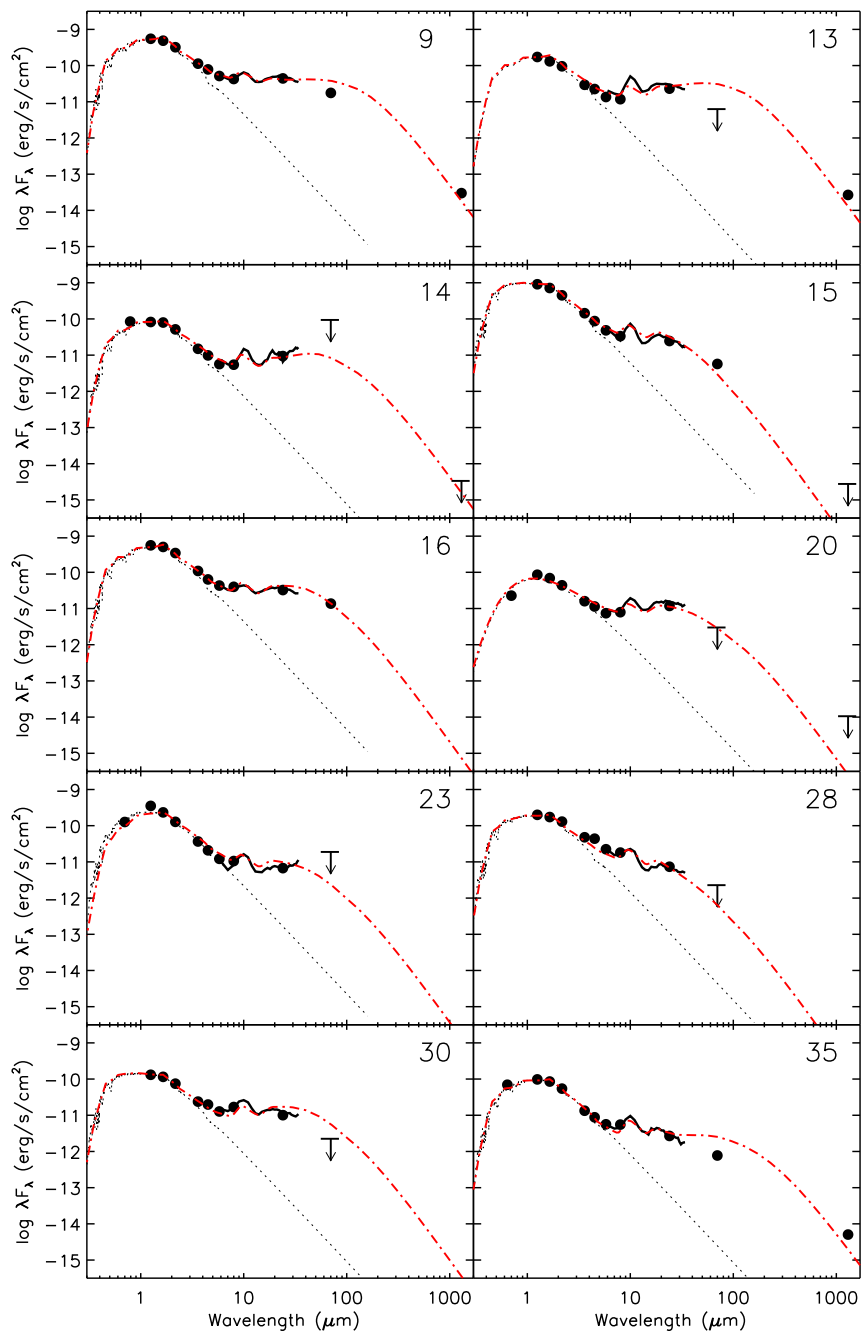}
\caption{Spectral Energy Distributions of the disks without holes in the
sample. Points are dereddened fluxes from IRAC, MIPS and IRAM. The
thick solid line is the IRS spectrum, the dotted line is the stellar photospheric model
and the red dash-dot line is the best-fit RADMC disk model. The numbers
give their identifications in Table \ref{tab:sample}. \label{seds_figure_noholes}}
\end{figure}

In order to determine the presence or absence of a hole, the full SED
is needed to determine the disk structure. The SEDs for the stars in
the sample are organized by type with edge-on disks (Figure
\ref{seds_figure_edgeon}), disks with holes (Figure
\ref{seds_figure_holes} and \ref{seds_figure_holes2}), and disks without holes (Figure
\ref{seds_figure_noholes}). The solid dots are dereddened fluxes,
which contains the 2MASS JHK fluxes, the 4 IRAC fluxes from 3.6 to 8
$\mu$m and the MIPS fluxes at 24 and 70 $\mu$m for all sources, plus
optical or 1.3mm fluxes for a smaller subsample of the objects. In
some cases, 70 \micron\, photometry produced only upper limits, often
accompanied by non-detections at 1.3mm, and thus severely limits
information about the outer disk. The solid dots are the dereddened
fluxes, the solid line is the IRS spectrum binned at a resolution of 1
\micron, the grey dotted line is the stellar Kurucz model of the
central star. The differences between the Kurucz and the NEXTGEN stellar 
models used above are negligible for the purposes of disk characterization.

The data were dereddened using the $A_{\rm V}$ values listed in Table
\ref{tab:stellarparameters} and the extinction law by
\citet{Weingartner2001} with $R_V$=5.5. Recently, \citet{Evans2009}
and \citet{Chapman2009} have suggested that this larger value of $R_V$
is more suitable for star forming clouds than the typical value of
$R_V$=3.1, which is based on models and observations of the diffuse
interstellar medium. The higher $R_V$ value reflects larger maximum
dust grain sizes and results in stronger dereddening effects
throughout the mid-IR. Dereddened fluxes are 2-4 times higher at 24
\micron\, and $A_V>$10 than with $R_V$=3.1. However, observational
measures of extinction indicate that the large amount of reddening is
in agreement with observations and may even underestimate the actual
effect in star forming regions (\citealt{Chapman2009, Flaherty2007}).

As the extinction law is not flat throughout the mid-IR, errors in the
mid-IR extinction law could potentially artificially create or remove
SED deficits mimicking inner holes. Corrections are much smaller at
low $A_V$, so effects only become significant for $A_V >$ 10. Based on
observations, the extinction law used here is conservative with some
cold disks being potentially missed due to inadequate dereddening
around 20 \micron .  With such large grain sizes in the absorbing
dust, silicate grain opacities are expected to have strong effects on
the 10 \micron\, silicate feature. In a few dereddened SEDs, strong silicate
features are seen which are not present in the uncorrected
spectra. The effects of extinction on the silicate features are
complex and dependent on extinction and line-of-sight ice and dust
composition, which may not be uniform throughout star forming regions
\citep{Mcclure2009}.  Due to this uncertainty, all work on the
silicate feature in this paper has been done on spectra which have not been dereddened,
with the caveat that sources with high $A_V$ have large errors (see also discussion in \citealt{Oliveira2010}).

The 2-D radiative transfer model RADMC \citep{Dullemond2004}, as
modified to include a density reduction simulating a hole, was used to
model the SEDs. The model has a large number of input parameters and
we have fixed as many as possible. The model assumes a passive disk,
which merely reprocesses the stellar radiation field and does not
account for accretion luminosity and heating. This is justified since
the disk to stellar luminosity ratios of the majority of the disks in
this study are below 0.4 (see Table \ref{tab:diskparams}). Further
discussion on the disk model dependency of the results can be found in
\S~\ref{colddiskproperties}. Stellar input parameters are mass,
temperature and luminosity which were determined from optical
spectroscopy and photospheric SED fits (see Table
\ref{tab:stellarparameters}). A stellar Kurucz model is then used for
the stellar radiation field. In terms of disk structure, the disk is
assumed to be flared such that the surface height, $H$, varies with
radius, $R$, as $H/R \propto R^{2/7}$, as in \citet{Chiang1997}. The
models do not calculate the hydrostatic equilibrium self-consistently
allowing the dust distribution to differ from the gas distribution
(expected to be in hydrostatic equilibrium) due to effects such as
dust settling. The pressure scale height is anchored at the outer disk
edge, in this case 200 AU. In some cases, some degree of settling
appears to have taken place. We present this number as a fraction of
the scale height predicted from hydrostatic equilibrium, assuming a
midplane temperature of 20 K. In cases where the disk has settled, a
flaring index of 2/7 (0.29) is likely too large (see
e.g. \citealt{Andrews2009} where flaring indices range from 0.04 to
0.26). Disk model SEDs with large flaring indices are similar to cold
disk SEDs in having more flux at longer wavelengths. By using a large flaring index as our default,
we are more conservative when reporting the presence of a
hole. Surface density is a power law of radius with an index of -1
inside 200 AU and then steepens to an index of -12 outside the 200 AU
outer disk edge. The dust composition is set to a silicate:carbon
ratio of 3:1 with only amorphous, as opposed to crystalline, silicate
included. Silicate opacities were taken from \citet{Beckwith1990}
which provides opacities into the millimeter. Disk masses were taken
from 1.3mm observations where possible (see Table
\ref{tab:diskparams}) and left as a free parameter otherwise. Disk
mass and disk settling are highly degenerate parameters in the absence
of (sub)millimeter fluxes, so when the disk mass is unknown, the scale
height is left at the hydrostatic value (see note on exception for
sources 18, 19 and 24 in Section \ref{individualnotes}). As a result,
model-derived disk masses tend to be systematically small and together
with observational upper limits provide a range of likely disk
masses. The inner edge of the disk, when no hole is present, was set
at approximately the radius where the dust sublimation temperature,
1500 K, is reached. Some cold disks in the literature have
near-infrared excesses most easily modelled as a gap rather than a
hole (e.g. \citealt{Brown2007,Espaillat2008}). However, none of the
sources in this sample required warm dust within the hole. Those
sources which did have some near-infrared excess turned out to be
compatible with disk models without holes. The hole is represented in
the model by a hole radius, $R_{\rm Hole}$ and a density reduction,
set here to a factor of 10$^{-6}$, providing an almost entirely clean
hole. Thus, the four free parameters are $R_{\rm Hole}$, degree of
settling, disk inclination and disk mass when there are no
measurements.

A two step procedure was used to find the best model fits. First, we
ran a large grid of disk models with a broad range of disk and hole
parameters and a limited number of stellar templates. Then, we ran a
finer grid of disk and hole parameters, determined from the larger
grid, for each individual star using the stellar properties in Table
\ref{tab:stellarparameters}. A simple $\chi^2$ minimization was
performed at each step between the dereddened SEDs and the model SEDs
and the resulting best fit is shown as a red dash-dot line in the SED plots
(Figures \ref{seds_figure_holes} and \ref{seds_figure_noholes}). Errors
on the hole radii are given in Table \ref{tab:diskparams} based on fits with up 
to 20\% deviation in $\chi^2$.

\begin{figure}
\includegraphics[scale=0.5,angle=90]{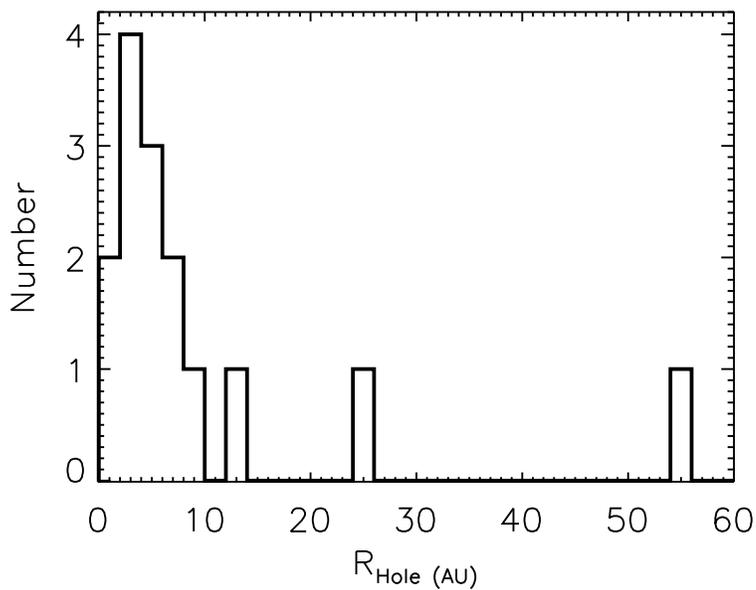}
\caption{Inner disk radius histograms for the cold disks in the sample. The cold disks
predominantly have inner holes with radii smaller than 10 AU. Many of the cold disks in the literature have larger hole sizes.}
\label{fig:rholehist}
\end{figure}

Fifteen of the 25 modelled stars require inner holes while the
remaining 10 require no hole to fit the SED. The 10 sources not
modelled are the 8 edge-on disks and 2 sources, 25 and 27, which have
strong extended 70 \micron\, emission and cannot be fit with disk
models (see section \ref{individualnotes} for further details). Hole
radii range from 1 AU to 55 AU. Holes smaller than 1 AU cannot be
firmly identified. The range of hole radii can be seen in histogram
form in Figure \ref{fig:rholehist}.

In order to confirm our results above and identify the less model-dependent 
part of our results, we also fitted our sample SEDs with the
on-line SED fitting tool by \cite{Robitaille2007}. This system offers
the possibility of fitting young star SEDs with a precomputed grid of
200,000 synthetic SEDs from stars, disks and envelopes with a broad
range of physical parameters. In order to make these fits, we 
observed photometry from Table \ref{tab:phottable2} and \ref{tab:phottable1} 
along with a binned IRS spectrum in from 6 to 33 $\mu$m with steps
of 1 $\mu$m. An aperture of 4.5 arcsec, a distance range from 0.1 
to 0.33 kpc and a A$_V$ range from 0 to 30 mags were used in all
fits. For the objects for which the spectral types are known, we select
the best fitting model with an effective temperature less than 500 K away
from the target's one. The final best-fit models used are
good representation of the median properties of the best fitting models 
with the appropriate effective temperatures but should not be considered
as definitively constrained parameters.

Figures \ref{seds_figure_holes} and
\ref{seds_figure_holes2} show the Robitaille et al.  best-fit models
for the objects in our sample as thin black dashed lines and
Table~\ref{tab:diskparams} shows their corresponding model ID's and
the inner disk radii in the cases where it is clearly distinguishable from the
dust destruction radius disk inner radius. In order to overplot the Robitaille et al. 
models, we corrected them with the extinction law given in the web-fitting
tool and using the total A$_V$ value given by the best-fit model and normalized
it to the $K_{\rm 2MASS}$-band dereddened flux. The $A_V$ values of 
the best-fit models used are in general in agreement with the $A_V$ values obtained from 
the spectral type and optical photometry. However, small discrepancies plus small differences 
in the standard extinction laws used to deredden our photometry and the Robitaille
models can explain the discrepancies between the Robitaille models and optical fluxes 
in some cases.

All but source 18 of the cold disks in Figure \ref{seds_figure_holes} also have inner holes with 
the Robitaille et al. models. The only exception, source 18, is a border-line case with
an extremely small inner cavity of 0.03 AU in the Robitaille et
al. best-fit model. On the other hand, almost all the objects in
Figure ~\ref{seds_figure_noholes}, which have no hole according to the
RADMC models, show inner cavities smaller than 2 AU in the Robitaille
models.  The only exception is source 15, which is also among the
flattest disk in the group.  This indicates that in general terms, the
classification of objects as disks with or without inner holes is
model independent as long as the disk hole is larger than 2 AU.

\begin{figure}
\includegraphics[scale=0.7]{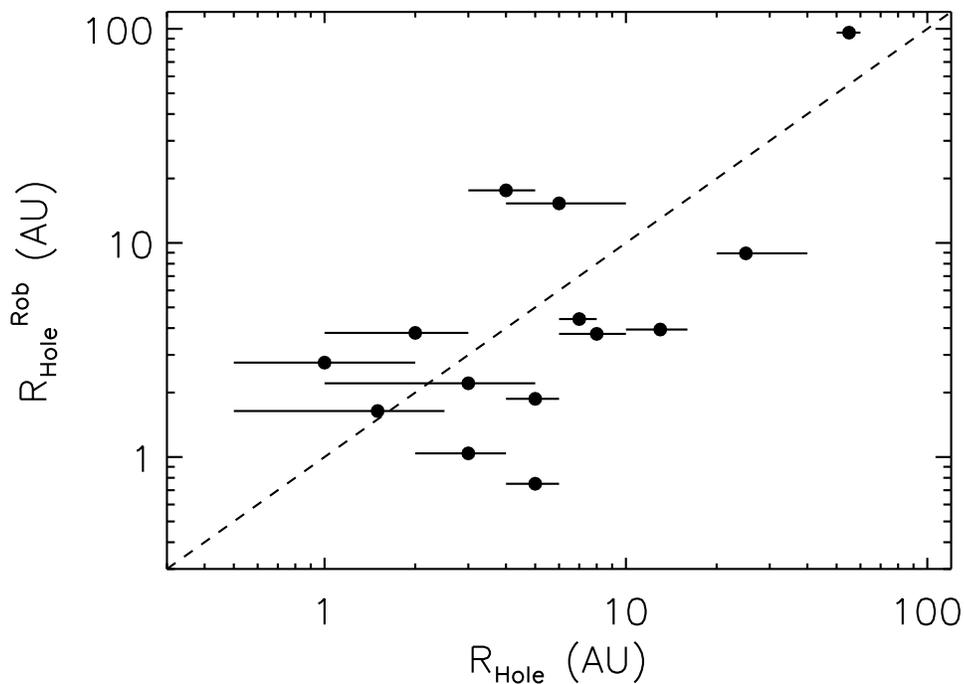}
\caption{Comparison between hole radii computed with the RADMC models
  \citep{Dullemond2004} and with the \cite{Robitaille2007} disk model
  grid. Even with the large scatter, both quantities are correlated,
  which shows that the derived disk hole radius is model independent
  to generally within a factor of 2-3.}
\label{fig:rhole_comparison}
\end{figure}

Figure \ref{fig:rhole_comparison} compares the disk inner holes
estimated both with the RADMC and the \cite{Robitaille2007} disk model
grids. The plot does not include the only outlier (Source 18). For the
rest of the sample, the values from both sets of models are
correlated, but with a very large scatter. It must be noted that this
comparison is not straight forward as both sets of models have
different treatments of the dust opacities, disk flaring angle, and
inner hole structure. Also, the Robitaille et al. on-line tool
does not allow the use of a priori information like the lack of an
envelope, or the effective temperature of the central star, both of
which have obviously great impact on the final result of the
fitting. Pursuing a more detailed comparison between both sets of disk
models is beyond the scope of this paper but this comparison shows that the derived disk hole
radius is model independent to within a factor of 2-3 generally. We find
the RADMC models more reliable due to the greater control over the input
parameters and the resulting better fits. The RADMC disk radii are
therefore used in all further comparisons.

\subsection{Dust composition}
\label{mineralogy}

A wide variety of mid-infrared spectral features have been detected in
spectra of disks around low- and intermediate mass young stars using
ISO (e.g., \citealt{Acke2004}) and {\it Spitzer} (e.g.,
\citealt{Kessler-Silacci2006, Furlan2006, Lahuis2007, Bouwman2008,
  Watson2009, Olofsson2009}). These features can serve as diagnostics
  of physical processes in disks such as grain growth,
  fragmentation, crystallization, flaring, and UV or X-ray illumination. An
  interesting question is whether the spectral features of cold disks
  differ significantly from those of normal T Tauri disks, and if so,
  what that tells us about the disk structure and evolution.

Table \ref{tab:sampleclass} identifies the presence or absence of the
10 and 20 $\mu$m silicate emission features, PAHs and crystalline
silicate features, where "Y" means a detection, "N" a
non-detection and "T" a tentative detection. The spectra presented in 
Figures \ref{IRS_edgeon} to \ref{IRS_CD_noacc} clearly show a large 
variety of features in our sample of cold disks.

\subsubsection{Silicates}

\begin{figure}[h!]
\includegraphics[width=\textwidth]{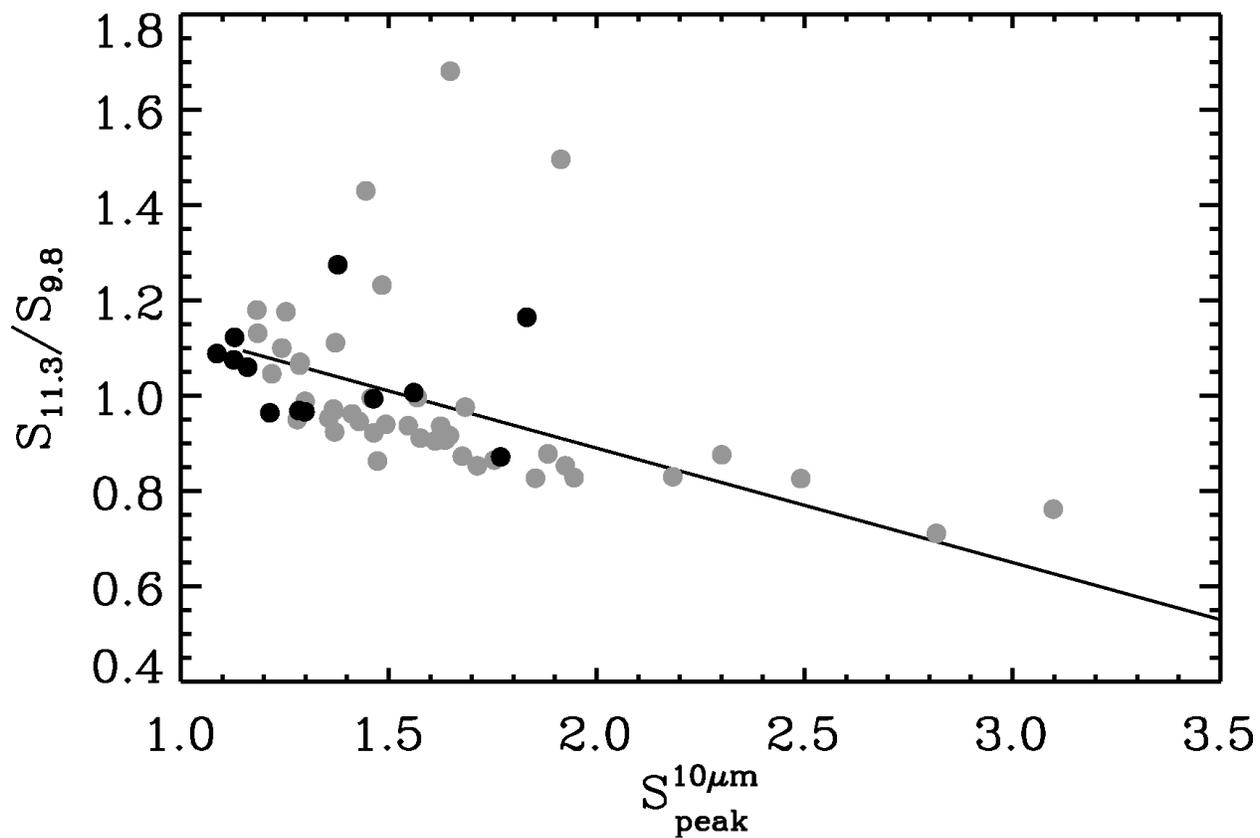}
\caption{Shape-strength diagram for the 10 $\mu$m silicate emission feature for all
cold disks in this sample (filled dots), compared with those of the c2d IRS T Tauri sample (\citealt{Olofsson2009}, grey dots). 
\label{Silicateshape}}
\end{figure}

The silicate 10 and 20 $\mu$m features are detected in most cases;
none of the sources show the featureless spectra commonly seen for
older (debris) disks (e.g., \citealt{Jura2004,Chen2006,
Carpenter2009}). However, only a few sources have pristine
ISM-like 10 $\mu$m silicate features, characterized by a strongly peaked 
profile centered at 9.8 $\mu$m. This is illustrated in Figure
\ref{Silicateshape}, in which the $S_{11.3}/S_{9.8}$ $\mu$m flux ratio is
plotted versus the strength of the feature above the continuum 
$S^{10 \mu m}_{\rm peak}$, as defined in \citet{Kessler-Silacci2006}. The 
different quantities $S_{11.3}$, $S_{9.8}$ and $S_{\mathrm{Peak}}^{10 \mu m}$ 
are calculated after normalizing the observed feature as follows 
$S_{\nu} = 1 + (F_{\nu} - F_{\nu, \mathrm{cont}}) / (<F_{\nu, \mathrm{cont}}>)$, 
where $F_{\nu, \mathrm{cont}}$ is the local continuum and 
$<F_{\nu, \mathrm{cont}}>$ the mean value of this continuum.
The local continuum is estimated using a second degree
polynomial, normalized at 6.8--7.5 $\mu$m, 12.5--13.5 $\mu$m and
30--36 $\mu$m. The fluxes are measured as an averaged flux of the
continuum-subtracted spectra in a $\pm$0.1 $\mu$m interval at each
wavelength. 

As shown by \citet{vanBoekel2005} and \citet{Olofsson2009}, sources in 
which the grain size distribution is dominated by
large $\mu$m-sized, or sources with size distributions much flatter than 
the MRN size distribution,  
grains appear in the upper left corner of the plot
whereas those dominated by smaller ISM-type grains are located in the lower-right
part. For comparison, the silicate profile parameters of the c2d
sample studied by \citet{Kessler-Silacci2006} and \citet{Olofsson2009}
are overplotted (grey dots). It is seen that the cold disks fall
mostly in the upper left part of the figure, indicating that the dust
in these disks has already undergone non-negligible grain growth. No
significant difference was found in the 10 $\mu$m shape for accreting
or non-accreting objects. This conclusion is not affected by
crystallization or sedimentation of dust toward the midplane, which
causes some spread in the relation but cannot reproduce the observed
trend, as discussed extensively in \citet{Dullemond2008} and
\citet{Olofsson2009}.

Some confirmed disks with inner holes show 
crystalline silicate features, either the forsterite feature at 33
$\mu$m and/or the 23 and 28 $\mu$m complexes
(see \citealt{Olofsson2009} for definition of
these features).  The fraction is between 33\% and 60\% (5-9/15) with the range due 
to some tentative detections limited by the low signal-to-noise ratio in some spectra. 
A prominent example of a highly crystalline spectrum
is that of object 10, SSTc2d J034227.1+314433, a cold disk in Perseus (see Figure
\ref{IRS_CD_noacc}). This detection frequency of crystalline material
is comparable to the 55\% found for the c2d sample of normal CTTS disks based on 
long-wavelength features, whereas \citet{Watson2009} claim an even higher fraction 
of 94\% for normal disks in Taurus (see discussion in Olofsson et al.\ 2009). 

\subsubsection{PAHs}
\label{pah}

\begin{figure}
\includegraphics[width=\textwidth]{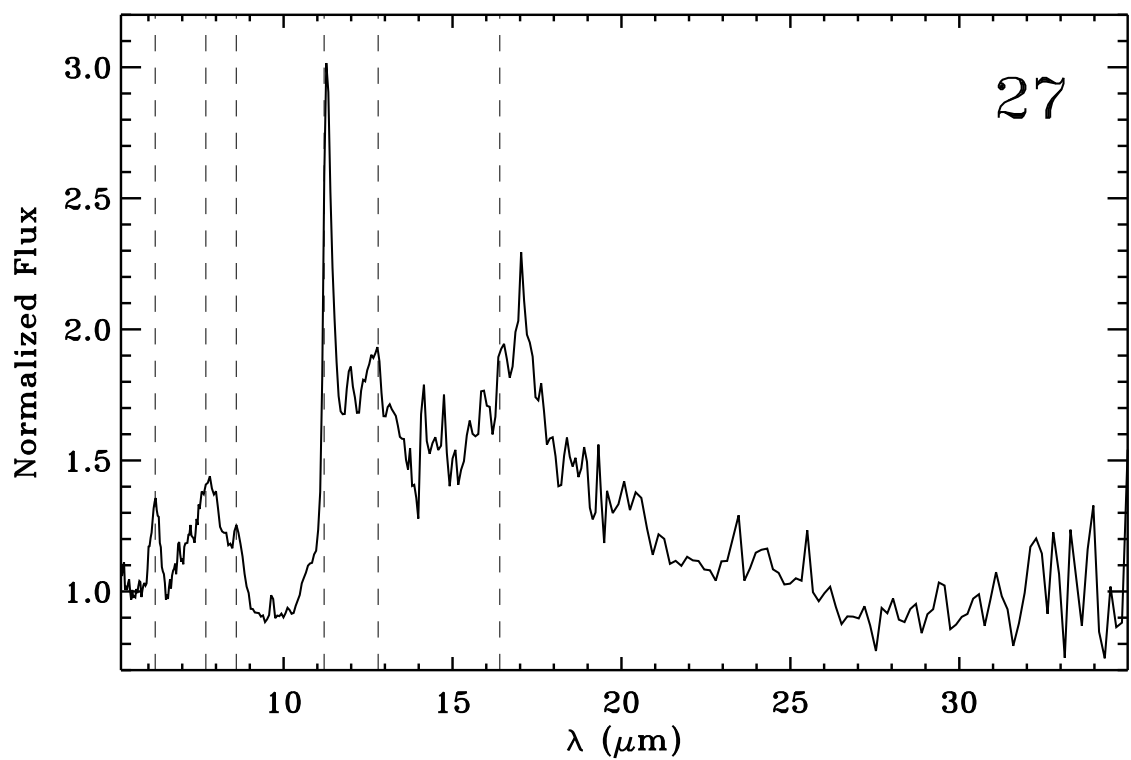}
\caption{Continuum subtracted PAH emission from object \# 27. The vertical dashed lines show PAH features.}
\label{fig:pah}
\end{figure}

The detection of PAHs is based on at least two of the features at 6.2,
7.7, 8.6, 11.2, 12.8 and 16.4 $\mu$m
\citep{Koike1993,Geers2006,Tielens2008}. Overall, PAH features are
detected in 13\% (2/15) of the cold disks in our sample. This is
similar to the fraction of 11--14\% found by \citet{Geers2006} for a
sample of normal CTTS disks (see also \citealt{Geers2007b}). This low
detection fraction not only reflects the fact that late-type stars
have less UV and optical photons to excite the PAHs but also implies a
PAH abundance in disks that is typically a factor of 10--100 lower
than in the normal interstellar medium, taken to be $5 \times 10^{-7}$
with respect to total hydrogen for PAH species with 100 carbon atoms
(e.g., C$_{100}$H$_{24}$) \citep{Geers2006}. Both sources 25 and 27,
which have long wavelength fluxes composed of extended material, show
exceptionally strong PAH features. Both have low continuum at
mid-infrared wavelengths increasing the line-to-continuum contrast. In
the case of source 27, even the 16.4 $\mu$m band is clearly revealed
(see Fig. \ref{fig:pah}).

The strong features seen in cold disks do not
necessarily imply higher PAH abundances than in normal CTTS disks,
however. \citet{Dullemond2007} model the PAH emission in disks
with varying degrees of sedimentation, as characterized by the
$\alpha$ parameter of turbulence. In the case of low turbulence, $\alpha
<10^{-4}$, the big grains quickly sediment to the midplane but the
PAHs stay in the upper layers, boosting the feature-to-continuum ratio
by a factor of 2--10. For high turbulence, $\alpha \approx 10^{-2}$,
the small and big grains stay well mixed in the upper layers,
resulting in the typically weak or absent PAHs features consistent
with observations \citep{Geers2006}. \citet{Dullemond2007} did not
consider disks with inner holes, but qualitatively the effect is
similar to their low turbulence case in which the big grains are
removed and the features strengths are boosted.

In contrast with the silicates, the PAH emission arises not only from
the inner disk but also from the outer disk surface exposed to optical
and UV radiation. Indeed, spatially extended PAH emission has been
detected out to $\sim$60 AU or more for some disks
(\citealt{Habart2006}, \citealt{Geers2007}). Two cold disks within the
c2d clouds, SR 21 and T Cha, are included in the observations of
\citet{Geers2007b} using VLT-ISAAC and VLT-VISIR to get higher spatial
resolution compared with {\it Spitzer}. In both cases, the emission is
found to be spatially unresolved, with limiting spatial extents (FWHM)
of 19 and 13 AU, respectively. These limits are comparable to the
values of $R_{\rm Hole}$ of 18 and 15 AU, respectively
(\citealt{Brown2007}, Table \ref{tab:diskparams}), indicating that the
PAHs are located primarily inside the gaps. For the case of SR 21, the
location of the gas inside the gap has been pinpointed to a ring at
$\sim 7$ AU radius \citep{Pontoppidan2008}, proving that the gas
emission is indeed coming from inside the gap and not from the outer
edge or wall of the gap. Since PAHs are likely coupled to the gas, PAH
emission, when present, serves as confirmation of the presence of gas
inside the gap. Unfortunately we see no correlation between PAHs and
accretion, both of which should trace gas within the holes. This is
likely due to the dependence of PAH emission on strong UV flux which
is lacking from late-type stars.

\subsection{Notes on individual sources}
\label{individualnotes}

Sources 10, 12, 14, 20, 22, 23, 29 and 30 have neither 70 \micron\,
nor 1.3 mm fluxes available. This makes it very difficult to
distinguish between a low mass disk and a more massive disk
with an inner hole. The difference between the two lies in the
presence of a substantial outer disk. The 70\micron\, limits sometimes
hint at the absence of such an outer disk but the limits are not
always very stringent. Unless there is compelling evidence from the
long wavelength IRS spectra, we have taken the conservative view that
these are likely low mass disks and can be explained without needing
to invoke an inner hole. 

Two sources, 25 and 27, have extremely strong, extended 70 \micron\,
emission and do not appear to be disks. In both cases, it is likely
that surrounding material has contaminated the long wavelength
photometry, producing these strange SEDs (see Figure
\ref{seds_figure_edgeon}).  Source 25, DoAr 21, has been better
studied than many of the other sources in this sample. Resolved
imaging of H$_2$ shows an extended ring structure at 73-219 AU away
from the star (Hogeheijde et al. 2009, submitted).
\citet{Jensen2009} have speculated that much of the flux in long
wavelength unresolved photometry comes from cloud material that the
star is heating. Thus, while the SED looks compatible with a large
inner hole, this might be due to excess material not directly
associated with the star with long wavelength fluxes being
increasingly affected.  Source 27 has a similar SED and we suspect may
arise from a similar physical situation. Both objects show
exceptionally strong PAH emission (see Figure \ref{fig:pah} and
\S~\ref{pah}). Within a disk structure, the strong 70\micron\, fluxes
indicate large amounts of dust at 100-200 AU and good SED fits
required unphysical input parameters. Objects with very pronounced 70
$\mu$m excesses but small 24 $\mu$m excesses have been found to be
produced by background contamination in the great majority of the
cases (Wahhaj et al. 2009, submitted). For these reasons, we have
removed these objects from further study.

The 70 \micron\, flux of source 17 is affected by strong variable
background emission, clearly visible in the c2d MIPS-70 image. Two
photometric points are given in the SED (Figure
\ref{seds_figure_holes}) with the larger being the full aperture flux
and the smaller a PSF aperture flux. The actual flux may be even
smaller. The 24 \micron\, image does not show background emission and
is in good agreement with IRS spectra. Although there is evidence for
background contamination in the long wavelength SED of this object,
a sensible fit was possible with a disk with a hole. 

Sources 11, 12, 13, 17, and 18 have stellar temperatures lower than the Kurucz
model grid and a blackbody is used instead.

Sources 18, 19 and 24 form a distinct subset of our sample. These
three disks all have little to no near-IR excess but very flat disks
longward of 8 \micron . While there does appear to be a discontinuity
between the inner and outer disks, the difference is not
large. Despite the unknown disk mass, some degree of settling in the
outer disk was required to reproduce the SED shape.

Finally, it is worth noticing that objects 2, 13 and 33 (15\% of the
cold disk sample) show factor of 2 differences between IRAC and IRS
fluxes in the overlapping wavelength range which are too large to be
due to calibration errors. In the particular case of object 33, which
has the largest difference, only the IRS spectrum hints at a possible
inner hole, while the IRAC photometry could be easily explained
without a hole. A careful check of the extraction of the IRS spectra
and IRAC photometry found no evidence of extended emission,
mis-alignment in the IRS observations, or any other instrumental
reason for the discrepancy. Indeed, only 6 objects (4\%) out of the
147 IRS observed disks in the Serpens IRS survey have similar flux
discrepancies between IRAC and IRS (Oliveira, I., 2010, priv. comm.). The
IRAC observations were taken approximately 2 years before the
spectroscopic ones (see Table \ref{tab:sample}). Similar scale
variability was reported in \citet{Muzerolle2009} for another
transitional disk. Such intrinsic variability might be the cause of
this phenomenon but systematics between spectra and photometry could
cause complications and further investigation is outside the scope of
this work.

\section{Discussion}
\label{discussion}

\subsection{Robust Selection Criteria}
\label{selection}
\subsubsection{Spectroscopic selection criteria for cold disks}

\begin{figure}[h!]
\includegraphics[scale=0.5,angle=90]{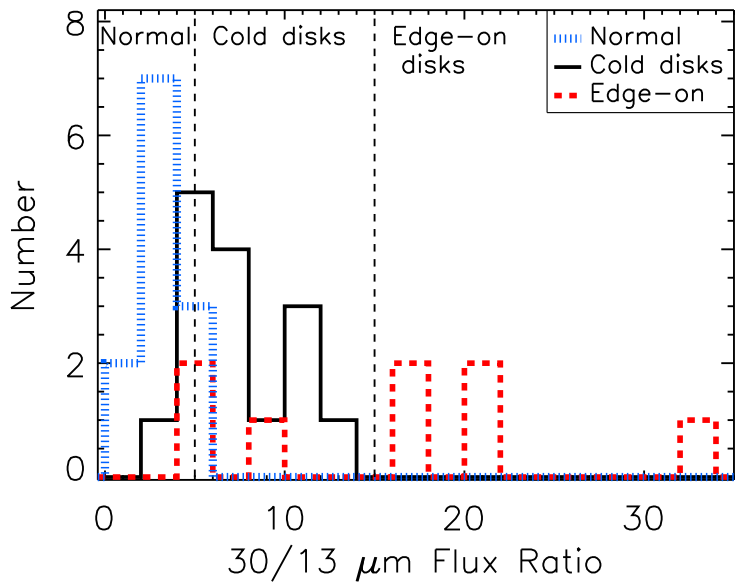}
\caption{Histogram of F$_{30}$/F$_{13}$ ratios for the disks in 
this paper, with the definition of the boundaries by \cite{Brown2007} marked by vertical dashed lines
for comparison. The region on the left is supposed to contain disks without holes, the middle region holds cold disks and on the right edge-on disks. The actual distribution of these types of sources in our sample is shown with normal disks in the blue short dashed line, cold disks with solid black and edge-on disks with the red long dashed line. There is significant overlap between all categories around F$_{30}$/F$_{13}$ of 5 but the general trends hold true in our sample.\label{fig:3013}}
\end{figure}

Determining concrete selection criteria for cold disks is necessary to
identify cold disks efficiently out of large {\it Spitzer} samples.
Previously cold disks have been identified in spectral samples using
30/13 \micron\, flux ratios and slopes
(e.g. \citealt{Brown2007,Furlan2009}).  These two wavelengths bracket
the rise in flux between an inner dust hole and substantial outer disk
for holes with radii between 1 and 100 AU around stars with a
representative range of stellar masses. The disks in this sample
provide confirmation of this classification with 93\% (14/15) of the
cold disks having $F_{30}$/$F_{13}$ flux ratios between 5 and 15 (see
Figure \ref{fig:3013}). The only disk with $F_{30}$/$F_{13}$$<$5 is
Source 2, which has one of the smallest holes. Edge-on disks generally
lie above $F_{30}$/$F_{13}$=15 but three overlap with the cold disk
region. There is also not a clean separation between normal and cold
disks at $F_{30}$/$F_{13}$=5. The 30/13 \micron\, flux ratio has a
crude correlation with larger hole sizes having larger flux ratios,
but there is unfortunately no clear relation between the 30/13
\micron\, flux ratios and the slope between {\it Spitzer} 8 and 24
\micron\, photometry, particularly for cold disks. Thus 8 to 24
\micron\, slopes and colors cannot be used as a proxy for estimating
hole size, although they can be used to select cold disks, as we
discuss in the next section.

\subsubsection{Photometric selection criteria}

To identify the cold disk candidates out of the YSO population, we
tested several combinations of color-color diagrams with different
{\it Spitzer} bands to determine the cleanest identification of disks
with inner holes (cold disks).  We included not
only the results of the detailed SED fitting from 
\S~\ref{diskparameters} but also literature information about cold disks
in the c2d clouds including two sources from \citet{Brown2007} and two
sources from \citet{Andrews2009}. The selection criteria proposed here
can be applied to any {\it Spitzer} IRAC and MIPS YSO photometric
sample of any star-forming region to identify the disks with inner
holes.

The criteria are based on colors without dereddening as the extinction
to many of the sources in the catalog is unknown. The effects of
A$_{\rm V}$=10 are plotted as an arrow on the diagrams. The effects
are small for extinctions less than 10. Analysis of the spectral
sample based on deredden colors provides little difference to the selection
regions.

\begin{figure}[h!]
\begin{minipage}{0.5\linewidth}
\hspace{-1cm}
\includegraphics[angle=90,scale=0.38]{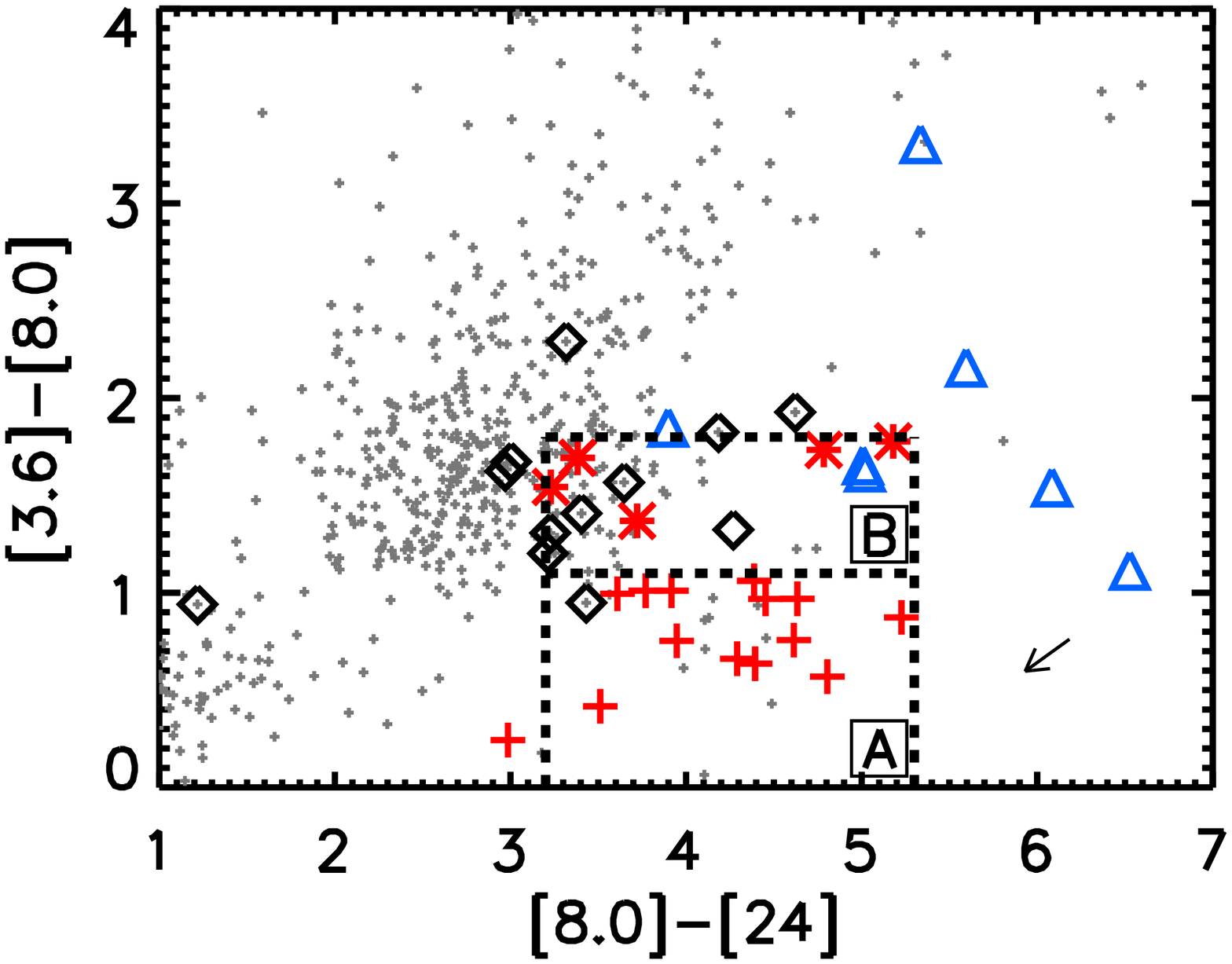}
\end{minipage}
\begin{minipage}{0.5\linewidth}
\hspace{-1cm}
\includegraphics[angle=90,scale=0.38]{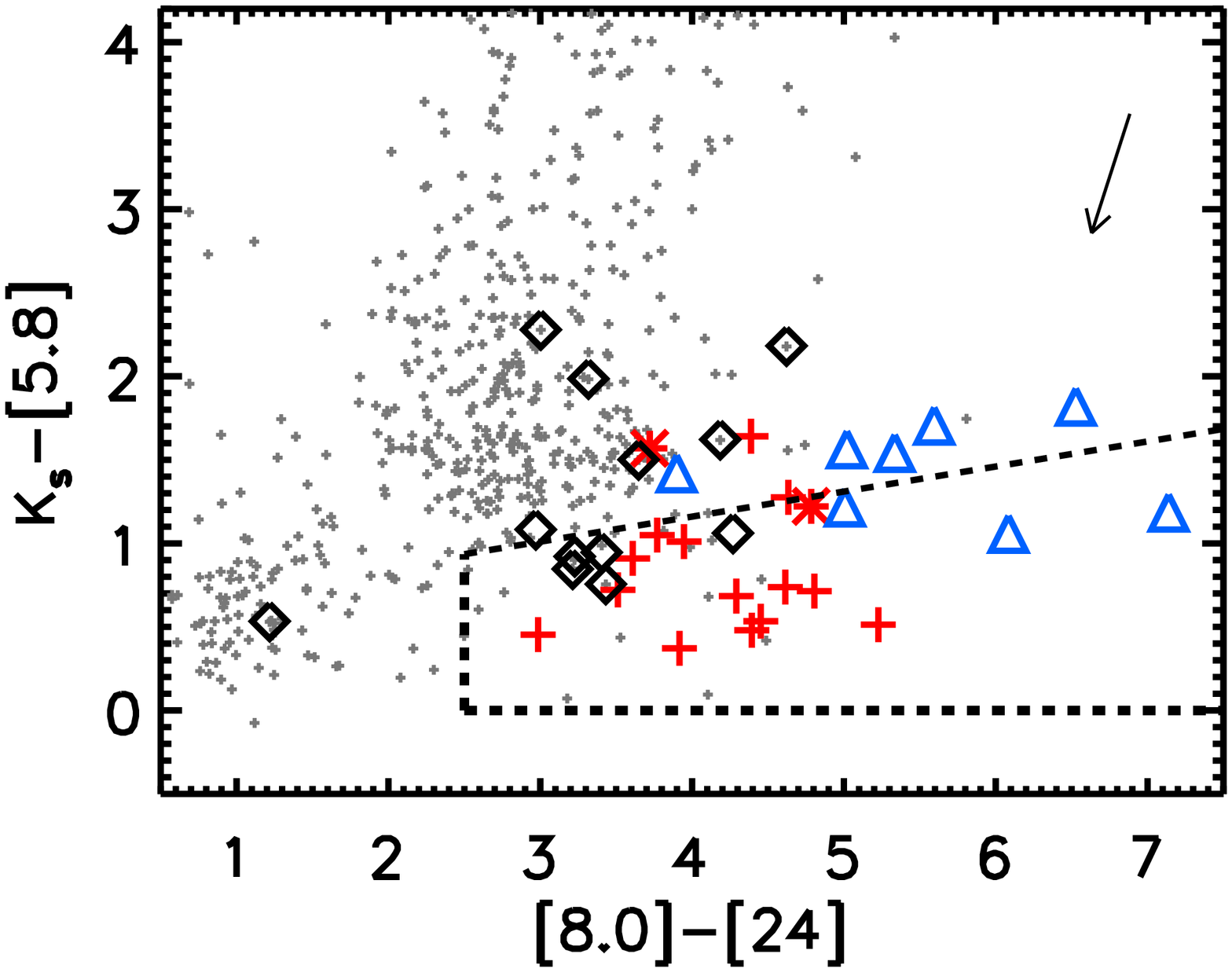}
\end{minipage}
\begin{minipage}{0.5\linewidth}
\vspace{-0.5cm}
\hspace{-1cm}
\includegraphics[angle=90,scale=0.38]{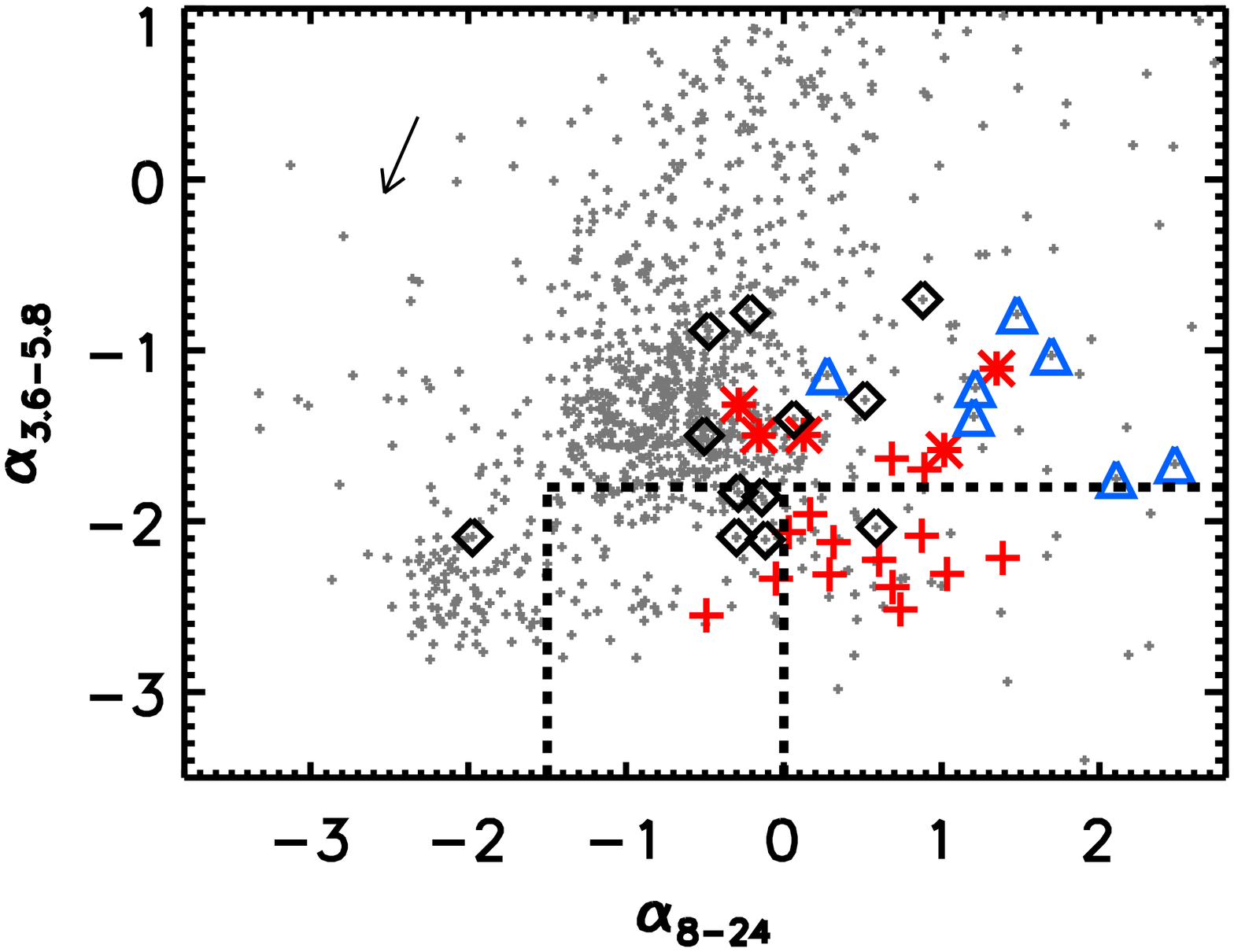}
\end{minipage}
\begin{minipage}{0.5\linewidth}
\vspace{-0.5cm}
\hspace{-1cm}
\includegraphics[angle=90,scale=0.38]{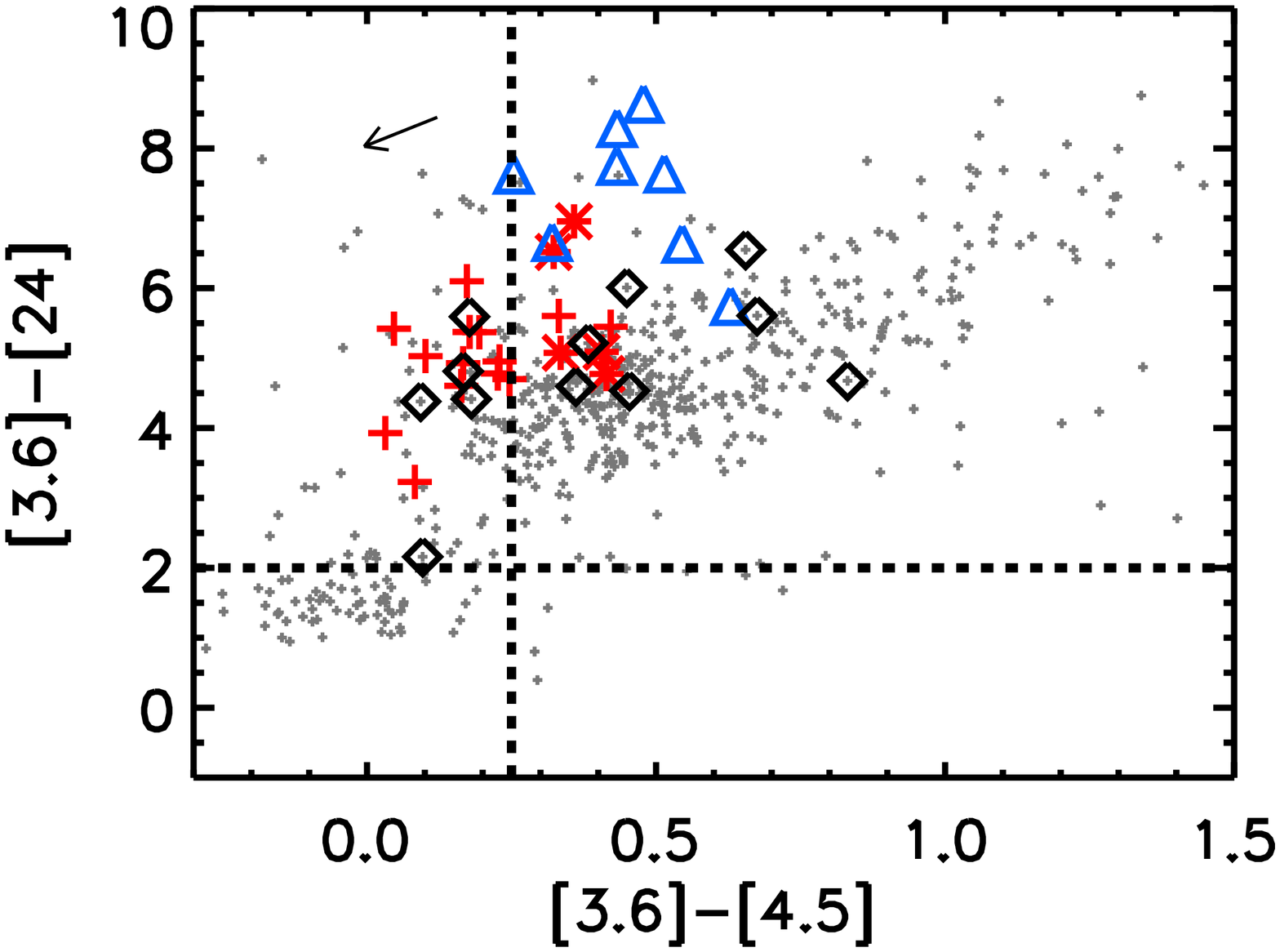}
\end{minipage}
\caption{Selection criteria. Red stars and crosses are the spectroscopically
  confirmed cold disks, black diamonds are disks without holes and
  blue triangles are edge-on disks. The grey crosses are all the
  sources in c2d YSO catalog \citep{Evans2009}. (top left) Our
  selection criteria. Regions A and B define the two selection criteria for identifying cold 
disks in photometric sample, the first one, represented with red crosses, selects "clean" inner holes and the second 
one, represented with red stars, identifies cold disks with near-IR excess. (top right) Selection criteria from
  \citet{Fang2009}. (bottom left) Selection criteria from
  \citet{Muzerolle2010}. (bottom right) Selection criteria from  \citet{Cieza2010} with transitional disks according to their definitions in the upper left quadrant. Arrows show the effect of correcting for an extinction of A$_{\rm V}$=10. 
\label{fig:criteria}}
\end{figure}

The cold disk candidate sample is defined with the following cuts in
magnitudes, organized in two sections for two different types of
objects (see top left Figure \ref{fig:criteria}): Region A selects
``clean'' inner holes (i.e. disks for which there is no substantial
excess in any IRAC band and the signature of an inner opacity
hole). Region B selects disks with some excess disk flux in the IRAC
bands. The definition of the boundaries are the following:

\begin{equation}
{\rm Region\,A:} \,\, 0.0\,<\,[3.6]-[8.0]\,<\,1.1\,;\,\,\,3.2\,<\,[8.0]-[24]\,<\,5.3
\end{equation}

\begin{equation}
{\rm Region\,B:} \,\,1.1\,<\,[3.6]-[8.0]\,<\,1.8\,;\,\,\,3.2\,<\,[8.0]-[24]\,<\,5.3
\end{equation}

\noindent Region B contains the 4 literature sources plus source 2,
while the majority of the new cold disks (13/15) lie in Region A. One
possible explanation for this division is the focus of this work on looking for
photospheric fluxes in the near-IR, while disks with large holes often
have small near-IR excesses (e.g. \citealt{Brown2007},
\citealt{Espaillat2007b}).

From the overall photometric catalog, the statistics on cold disks
within the five c2d clouds can be determined, including sources not
included in the spectroscopic follow-up for a variety of reasons such
as low fluxes. These disks are marked with small crosses in Figure
\ref{fig:criteria}.  There are 43 cold disk candidates in Region A
accounting for 4\% of the total disk population in the c2d YSO
catalog. Only one of the objects in that region for which we performed
detailed SED modeling including the IRS spectrum was a false
positive. This is object 25, DoAr 21, which has a cold disk-like SED
but the 70 \micron\, flux may be strongly contaminated by cloud
material (see \S~\ref{individualnotes} for further details). From the 
spectroscopically studied objects in Region A, 93\% are cold disks. 

Region B is more complicated. While there are well studied, confirmed
cold disks within this region, there are many objects which are not
cold disks. Region B contains 181 YSOs in the c2d catalog, but fewer
sources were selected for our study from this region. Of this small
subsample and excluding the rising sources, for which it is not
possible to determine the presence of an inner hole in the disk, 
only 42\% of the studied disks turned out to be cold disks,
so we estimate a contamination on the order of 58\% in this
region. These percentages have considerable uncertainty given
the low-number statistics. Region B includes the disks WSB 60 and DoAr 44, 
for which inner holes of 20 and 33 AU, respectively, have been resolved 
with the SMA but whose SEDs show no detectable signature of them 
\citep{Andrews2009}. This implies that selecting purely on broadband 
colors will always be prone to uncertainty. Individual source modeling, 
spectroscopic studies and non-SED based searches are all needed to 
fully understand cold disk frequencies, particularly in this region of 
the color-color parameter space.

If we add together the 43 objects in region A with 42\% of the 181
objects in region B, we have a total of 119 cold disks in the sample
(12\% of the YSO catalog of 1024 sources).  Therefore, the fraction of
4\% of cold disks quoted above should be interpreted as safe lower
limits to the actual population of disks with inner holes in the c2d
YSO catalog.  The analysis of this photometric sample is outside the
scope of this paper and will be presented in a separate
article. Considering the IRS flux-limited sample in Serpens where all
disks have IRS spectra, we confirm the presence of 8 cold disks, which
account for 9\% of the total \citep{Oliveira2010}. This range of cold
disk frequencies of 4-10\% comes with some caveats. Transitional disks
with very low mass outer disks are not included in this study (as the 
ones discussed in e.g. \citealt{Najita2007,Cieza2007}), but are
often included in other transitional disk frequency statistics. Such
disks may represent a later stage of cold disk evolution or a
completely different evolutionary pathway. Disks with holes smaller
than 1 AU cannot be reliably identified. Disks with large holes but
with some near-IR excess can also be difficult to identify from
photometry.

A final consideration is the potential bias introduced by the c2d YSO
selection criteria themselves. The c2d YSO sample is biased against objects with
photospheric colors and only includes sources with good detections at all IRAC bands
plus at MIPS 24 $\mu$m band. This obviously imposes bias which is a difficult to determine 
against selecting transitional disks in regions with very high background emission, 
like e.g. the IC 348 or $\rho$ Oph clusters. In any case, the effect seems 
not to be a dominant one since most of the cold disk candidates found with the color 
selection proposed here are clustered in a similar manner as the rest of the YSOs around 
the dense high-background clusters, while we should see a lack of such objects in high 
background areas otherwise.

\subsubsection{Comparison with transitional disk selection criteria from other papers}

\citet{Fang2009} also recently proposed color-color diagram cuts to
select cold disks. Their criteria form a trapezoid in K-[5.8]
vs. [8.0]-[24] color-color space defined by K-[5.8] $>$ 0, [8.0]-[24]
$>$ 2.5 and the last side by K-[5.8]$<$(0.56+([8.0]-[24])x0.15).  In
Figure \ref{fig:criteria}b, we evaluate these criteria based on our
spectrally confirmed cold disks. Most of the disks in our sample fall
within or close to this area. Of our cold disks, 13/15 are within the
boundaries while the remaining two are only just above the upper
sloped line. However, half (6/12) of the disks without holes also lie
within this region as well as some (3/8) of the edge-on disks. Cold
disks can be preferentially selected by cutting higher in [8.0]-[24]
at $\sim$ 3.5 instead of 2.5. This would exclude more of the
candidates that do not have inner holes while including all the disks
with holes and also have the benefit of having a greater separation
from the bulk sample in the larger catalog (shown in the figure as
small gray crosses). The region of [8.0]-[24] between 2.5 and 3.5 has
the potential to include transitional disks with low mass outer
regions, but individual disk modeling is needed to confirm the nature
of these sources. An upper limit in [8.0]-[24] is also needed to
remove the edge-on disks, with [8.0]-[24]$<$5 producing a reasonably
clean separation.

\citet{Muzerolle2010} use slopes in log $\lambda F_\lambda$ vs. log
$\lambda$ between 3.6 \micron\, and 5.8 \micron , $\alpha_{3-5}$, vs 8
\micron\, to 24\micron\,, $\alpha_{8-24}$ . They propose that all
transitional disks lie below -1.8 in $\alpha_{3-5}$ with weak excess
sources between $\alpha_{8-24}=-1.5$ and $\alpha_{8-24}=0$ and normal
cold disks at $\alpha_{8-24}>$0. Our sources lying within the weak
excess region were all found not to need holes to fit the SEDs. The
normal cold disk region selects similar sources to Region A but
sources within Region B are not selected by the \citet{Muzerolle2010}
criteria.

\citet{Cieza2010} adopt a wider definition of transitional disks
(e.g. they include objects with very small 24 \micron{} excesses), and
their selection criteria are therefore less useful for identifying
cold disks. Transitional disks are taken to be in the quadrant defined
by [3.6]-[24]$>$2 and [3.6]-[4.5]$<$0.25. Just over half (12/19) of
the cold disks lie within the selected quadrant, with 7 cold disks
missed by this criteria (generally those with more massive outer
disks).  Almost half (5/12) of the disks without holes lie within the
region and in the exact same position as the disks with holes. These
disks may be homologously depleted transitional disks. A fairly large
fraction of the catalog classical disk locus lies within the selected
region and might lead to a large false detection rate of cold disks.

\subsection{Properties of the cold disks}
\label{colddiskproperties}

One of the most obvious characteristics of this cold disk sample is
the relatively small hole sizes (see distribution in Figure
\ref{fig:rholehist}). Many of the IRS selected cold disks in the
literature (e.g. \citealt{Calvet2005, Brown2007, Espaillat2007b,
  Kim2009}) have large holes with radii greater than 10 AU. Of the 15
cold disks examined here, only three have hole sizes larger than 10 AU
while an additional four large hole sources within the clouds are
identified in the literature \citep{Brown2007,Andrews2009}. This may
indicate that cold disks with very large holes ($>$10 AU) are
intrinsically rarer than disks with smaller holes. Large hole sizes
are easy to see from IRS spectra but may not be as obvious from
photometry. One complication is that disks with large inner holes
often have some near-IR excess making the disks difficult to identify
from IRAC colors. Another possible complication arises from the
different disk models and methods used to determine the inner disk
hole radii.  In particular, the possibility cannot be ruled out that
the RADMC models, which do not include viscous heating due to
accretion tend to produce smaller inner holes than other accreting
disk models used by e.g. \citet{Calvet2002}. On the other hand, the
hole radii determined with same models and methods by \cite{Brown2007}
were confirmed by the direct SMA observations of three disks by Brown
et al. (\citeyear{Brown2009}). Finally, the similarity in terms of
infrared SEDs of the two millimeter-discovered cold disks in Ophiuchus
\citep{Andrews2009} to classical disks means millimeter surveys, as
well as infrared, would be needed to ensure that all cold disks were
found and to robustly confirm this trend. The differences in the
definition of transitional disks between the works of \cite{Najita2007} and 
\cite{Cieza2008} were also extensively discussed in \cite{Cieza2010}.

\begin{figure}
\includegraphics[angle=90,scale=0.7]{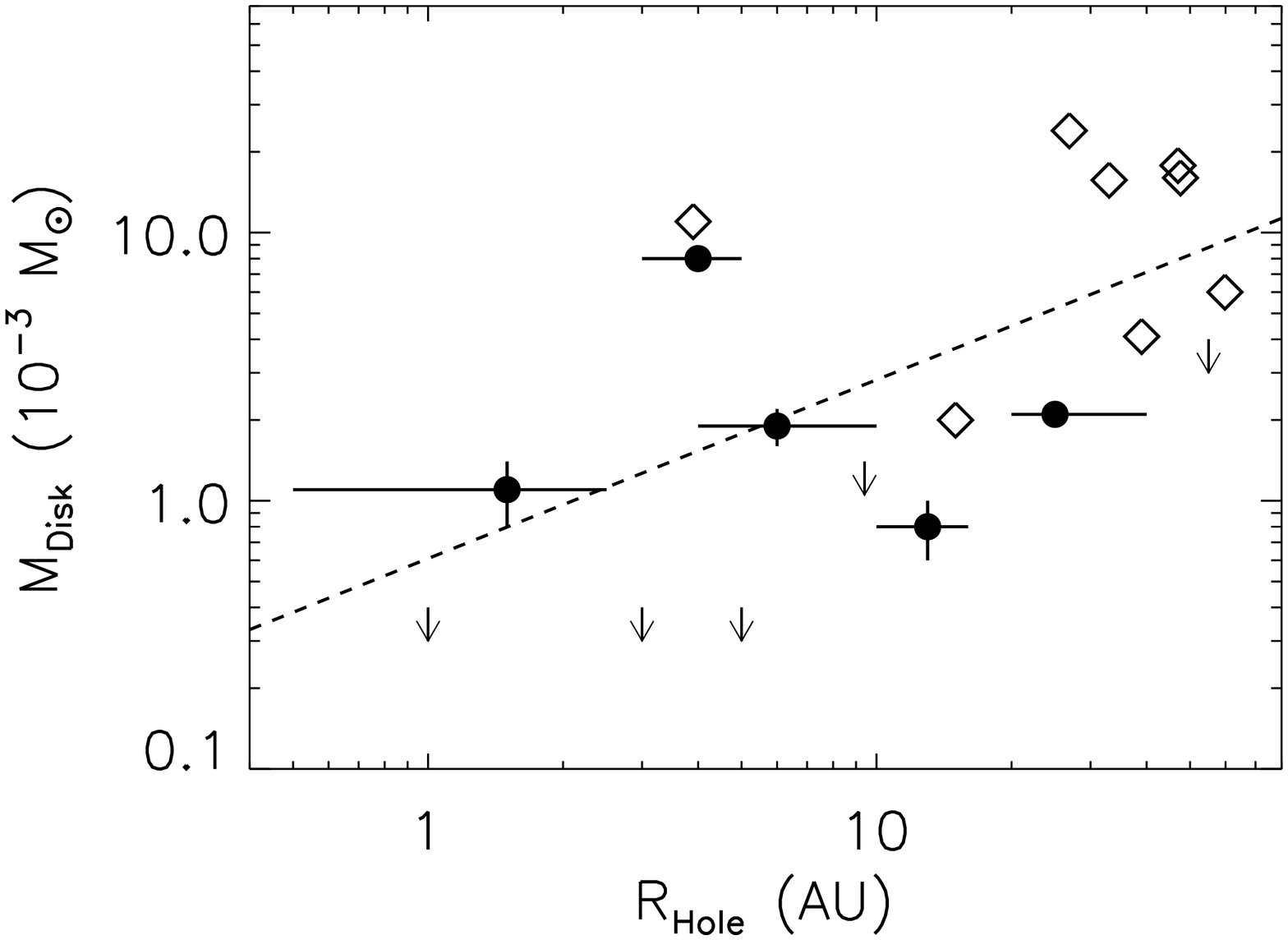}
\caption{Correlation between hole size and disk mass. Dots and upper
  limits represent source with millimeter fluxes from this work while
  the diamonds are the cold disks with millimeter measured disk masses
  from \citet{Kim2009} and \citet{Brown2007}. To minimize systematic
  differences, disk masses for sources from \citet{Kim2009} have been
  recalculated using the 1.3 mm fluxes in \citet{Andrews2005} and the
  conversion in Section \ref{diskmass}. The correlation has a
  Pearson's correlation coefficient of 0.6 and is statistically
  significant on the 99\% level including the upper limits and 90\%
  level with only the detected sources. The systematically lower disk
  masses in this survey might explain the generally smaller hole sizes
  than previously published transitional disks.}
\label{fig:dmass}
\end{figure}

We find a statistically significant positive correlation between disk mass
and hole size (see Figure \ref{fig:dmass}). The correlation has a
Pearson's correlation coefficient of 0.6 and is statistically
significant on the 99\% level including upper limits on disk mass and
on the 90\% level with only detected sources. To minimize systematic
differences, disk masses for sources from \citet{Kim2009} have been
recalculated using the 1.3 mm fluxes in \citet{Andrews2005} and the
flux to disk mass conversion in Section \ref{diskmass}. The dependence
of hole size on disk mass may point to a gravitational process, with
more massive disks more likely to form more massive planets in larger
orbits. The masses of transitional disks relative to
normal disks remains unclear. \cite{Najita2007} finds that
transitional disks have larger disk masses for a sample of
transitional disks in Taurus, although their disk classification
method is different. \cite{Cieza2008b} find very small disk masses
for a sample of non-accreting Weak-Lined T Tauri stars. This trend may
help explain the different results of the studies as \cite{Najita2007}
focused mainly on cold disks from the literature which generally have
large holes and are often accreting, while \cite{Cieza2008b} were likely more sensitive to
disks with smaller holes and no accretion.

\begin{figure} 
\includegraphics[height=\textwidth,angle=90]{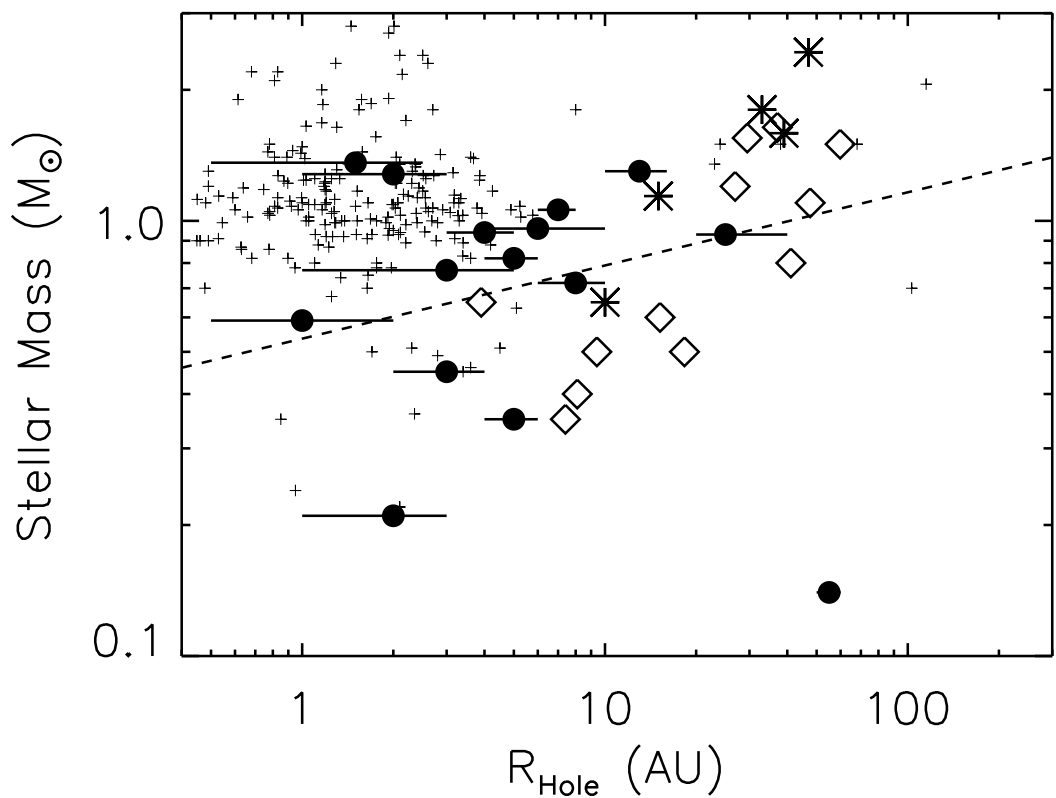}
\caption{Relationship between the size of the inner disk hole and the
  mass of the central star.  Solid circles represent our sample while
  stars and diamonds are transitional disks from \citet{Brown2007} and
  \citet{Kim2009} respectively. The dotted line is the best linear fit
  to the data. The small crosses mark the semimajor axis of all
  exoplanets in the {\tt http://exoplanet.eu} database as of December
  2009, compared to the mass of their parent stars. }
\label{fig:rholemstar}
\end{figure}

\citet{Kim2009} find a correlation between hole size and stellar mass
in their transitional disk sample in Chamaeleon and
Taurus. Unfortunately, our sample is too similar in stellar mass to
accurately test this trend and there is no correlation found from our sample alone. However,
combining with the \citet{Kim2009} sample (see Figure
\ref{fig:rholemstar}) retains a correlation between the size of the
inner holes and the stellar masses. The disks from \citet{Brown2007} are
in agreement with the \citet{Kim2009} trend while the disks in this
paper generally lie below the trend line, indicating smaller hole
sizes. This may be partly due to the generally lower disk masses 
in this sample leading to smaller hole sizes based on the correlation in Figure \ref{fig:dmass}.

Perhaps surprisingly, we see no trends with $L_{\rm Disk}/L_*$, a
measure of overall disk evolution (see Table \ref{tab:diskparams}).
$L_{\rm Disk}/L_*$ measures the integrated infrared excess normalized
by the stellar luminosity and primarily traces the total grain surface
area that is reprocessing stellar light. As disks become more 
tenuous, settle and disappear, the strength of the disk luminosity should
decline. This measure is commonly used in debris disk studies
(e.g. \citealt{Morales2009}). There is no correlation with hole size
nor are there significant differences in $L_{\rm Disk}/L_*$ between
the sources with and without holes in this sample. Therefore, if $L_{\rm Disk}/L_*$
can indeed be used as a probe for disk flaring, then this points to a
distinct lack of settling in the outer regions of cold disks. Alternatively,
this might also imply that $L_{\rm Disk}/L_*$ is not especially sensitive
to the changes in the disk flaring while the outer disk keeps being optically
thick.

The majority (9/12, 75\%) of the cold disks with H$\alpha$ spectra are
accreting. Accretion is a particularly interesting diagnostic in
transitional disks as it indicates material inside the hole. Large
fractions of accreting cold disks indicate that in many cases gas must
be flowing through the hole as any inner gas reservoir would drain
quickly. This is in contrast with the results by \cite{Najita2007},
who report lower mass accretion rates in transitional disks with
respect to the classical T Tauri stars in Taurus, although using a
completely different definition of transitional disk from the one used
here.

\begin{figure}[h!]
\includegraphics[height=\textwidth,angle=90]{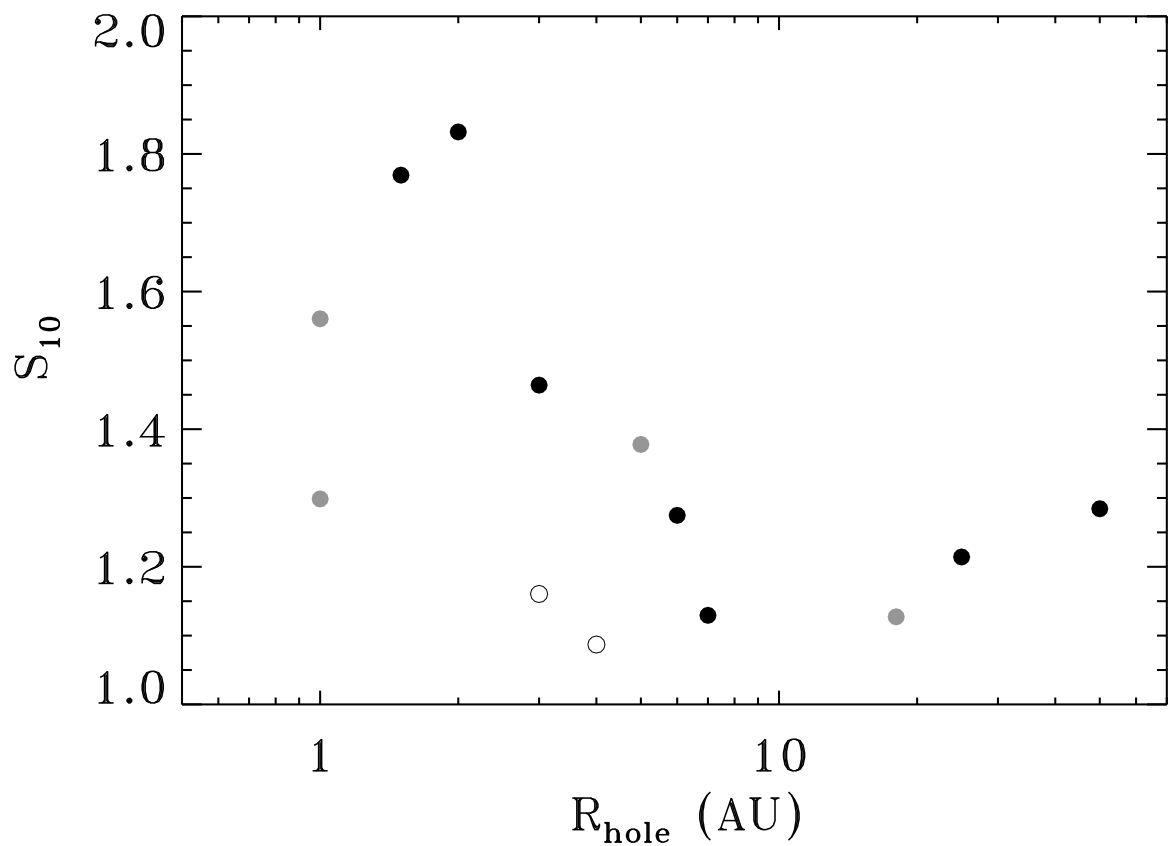}
\caption{Relationship between the size of the inner disk hole and the intensity of the 10 $\mu$m feature. 
Solid, grey and open circles represent c2d accreting, non-accreting cold disks and objects for which we
do not have yet accretion information, respectively. \label{fig:rhole_silicates}}
\end{figure}

Silicate emission probes dust properties such as grain size and
crystallinity. Silicate emission, especially the distinct 10
\micron\, silicate stretching band, is seen from all but one of the
cold disks. However, the presence and intensity of the 10 $\mu$m
feature decreases substantially for inner holes radii larger than 7-10
AU (Figure \ref{fig:rhole_silicates}). This is partly to be expected
as dust-depleted inner holes result in a lack of silicate grains at
the correct disk temperatures to produce the features.  A large
fraction of the cold disks show crystalline features at wavelengths 
longer than 20 \micron\, indicating processing which likely occurred 
in the inner regions of the disks where temperature and densities 
are high. These regions are now largely depleted in the cold disks 
but dust mixing, fragmentation of larger bodies or alternative formation 
routes have resulted in detectable amounts remaining further out in 
the disk.

\subsection{The possible origins of the inner holes}
\label{holeorigins}

The origins of inner holes are still under debate among several
theories developed to explain inner holes and gaps in protoplanetary
disks, including i) EUV photoevaporation of the inner disk
\citep{Clarke2001,Alexander2006}, ii) settling and coagulation of dust
into large particles \citep{Tanaka2005, Dullemond2005}, or iii) by the
dynamical perturbation of unresolved companions, both stars
\citep{Artymowicz1994} and planets
(\citealt{Quillen2004,Varniere2006b}).  It is likely that all are
present on some level but it is not clear which are dominant in which
systems \cite{Cieza2010,Sicilia-Aguilar2010}.

For photoevaporation, an inner hole occurs when the photoevaporation
rate driven by the EUV ionizing flux from the central star matches the
viscous accretion rate \citep{Alexander2006}. FUV photoevaporation
also plays a significant role in dissipating disks but predominantly
removes less bound gas from the outer regions and would therefore not
create cold disk SEDs \citep{Gorti2009}. Photoevaporation is most
effective when accretion rates are low and would result in no gas or
dust close to the star within 0.1-1 Myr, depending on assumed disk
properties such as viscosity and treatment of the UV ionizing flux
(\citealt{Alexander2006,Gorti2009}). The presence of accretion in such
a large fraction (75\%) of this sample discards photoevaporation as
the primary origin. Some disks with inner holes do show evidence for
photoevaporation based on [Ne II] observations but the implied mass
loss rates are too low to disperse the disk in under 10 Myr
\citep{Pascucci2009b}. The disk masses of the remaining 25\% are
typically low enough that photoevaporation could be the cause of the
inner hole.

Accelerated grain growth in the inner regions could produce cold disk
SEDs as dust grains became too big to be observed
\citep{Tanaka2005}. Models predict that dust particles grow and settle
towards the dense disk mid-plane, where they may stick together to
form planetesimals \citep{Weidenschilling2000}.  Growth is likely
preferential in the inner disk, and these larger bodies will grow and
can eventually accrete a large fraction of the surrounding gas to
become giant planets (\citealt{Bryden1999}; \citealt{Wuchterl2000} and
references therein).  In this scenario, grain growth and settling
would happen throughout the disk but with faster timescales in the
inner region. If grain growth is accompanied by settling throughout
the disk, decreasing $L_{\rm Disk}/L_*$ with increasing hole size
would be expected. We do not see this for our sample making this
unlikely to be the dominant scenario although such a correlation for a
subsample of the sources may be masked by short inner disk settling
timescales or objects formed by different mechanisms. Another
potential complication is that coagulation in the inner regions would
not effect $L_{\rm Disk}/L_*$ in the absence of settling in the outer
disk. Holes created via grain growth are likely to have more gradual
hole edges compared to holes created by companions.

Binary companions are a possible explanation for cold disk
SEDs. Some candidate cold disks, such as CoKu Tau 4, have been
later determined to be circumbinary \citep{Ireland2008}. The issue
clearly requires further high-resolution imaging of these sources
(e.g.  \citealt{Marois2008, Kalas2008,Lagrange2008}) or radial velocity
measurements to confirm the possible presence of stellar or
substellar-mass companions. However, this is time consuming for large
samples and may not even be possible for more distant regions or
planets at larger radii which require long term
monitoring. Preliminary results of binary searches show that close
binary companions often result in the complete dissipation of the disk
and that circumbinary ``transition'' disks may not be common
\citep{Kraus2009,Pott2010}. If stellar companions carve
out inner holes, accretion may either continue or cease
depending on the separation and mass ratio of the two stars
\citep{Artymowicz1996,Ireland2008}.  On the other hand, if a planet
carves out an inner hole accretion can continue at a reduced rate aided by 
mechanisms such as the magneto-rotational instability  \citep{Chiang2007}.

Inner holes created by planets remains a popular and exciting
explanation. Of the scenarios described this is the most difficult to
positively confirm or deny. Planets are faint so direct observation is
difficult while their small masses leave weaker gravitational
signatures on the disk than stellar companions. The inner hole radii
are compatible with the distribution of exoplanet semi-major axis, as
found in the latest exoplanet database as of December 2009\footnote{\tt
http://www.exoplanets.eu} (Fig.~\ref{fig:rholemstar}). As the exoplanets 
likely formed in similar disks before dissipation, it is likely that there are 
young exo-planets in some disks at these types of radii but whether they 
are responsible for the cold disk signatures discussed in this paper 
remains to be determined.

\section{Summary}
\label{conclusions}

The main results of this work are summarized here:

\begin{itemize}
\item Optical spectra, 2MASS and {\it Spitzer} photometry, millimeter
  continuum observations and {\it Spitzer}/IRS 5 to 35 $\mu$m spectra
  of a sample of 35 cold disk candidates selected from c2d photometry
  are presented and analyzed.
\item Out of 35 objects in the initial sample, SED modeling identifies
  15 as disks with inner holes, which we call ``cold disks'', following
  the c2d convention \citep{Brown2007}. Of the remaining sources, 10
  could be modelled without holes, 8 are edge-on disks and 2 have SEDs
  strongly contaminated by cloud material.
\item The color cuts $0.0<[3.6]-[8.0]<1.1$ and
  $3.2<[8.0]-[24]<5.3$ identify most cold disks from this
  sample ($\sim$ 80\%), in particular those with the cleanest inner
  holes. Extension of the $[3.6]-[8.0]$ color cut to 1.8 recovers some objects with
  small near-IR excesses and large holes, but contains 
  contamination from disks without holes. Out of the large c2d YSO sample,
  between $\sim$ 12\% of the disks are estimated to be cold disks
  based on these selection criteria.
\item We evaluated the criteria of \citet{Fang2009},
  \citet{Muzerolle2010} and \citet{Cieza2010} and suggest
  improvements based on our spectroscopic study.
\item The cold disks presented here have small hole sizes, generally
  less than 10 AU. This distribution is more in agreement with
  exoplanet radii than the large hole sizes of most cold disks in the
  literature.
\item A large fraction (75\%) of the cold disks are accreting,
  suggesting that gas is flowing through the dust depleted hole. This large
  fraction of accreting disks is not in agreement with the dominant
  hole origin being photoevaporation.
\item Hole size correlates with disk mass with more massive disks tending to have larger holes.   
\item The sizes of the inner holes scale linearly with the stellar mass although with a large
spread.
\item The 10 $\mu$m silicate features in the sample show substantial
  grain growth. The 10 $\mu$m silicate emission
  feature strength with respect to continuum decreases drastically for 
  inner holes larger than $\sim$ 7 AU. Some (33-60\%) of the cold disks show 
  long wavelength crystalline features
  indicating that mixing from the inner regions where crystallization
  occurs to outside the inner hole region must be efficient. Only 2 sources
   ($\sim$ 13\%) show PAH emission.
\end{itemize}

\acknowledgments

Support for this work, part of the {\it Spitzer} Space Telescope Legacy
Science Program, was provided by NASA through Contract Numbers
1224608 for c2d, 1288664 for GO3 IRS cold disks, 1256316 and 1230780 issued by the Jet Propulsion Laboratory, and for
California Institute of Technology under NASA contract
1407. Astrochemistry at Leiden is supported by a NWO Spinoza and NOVA
grant, and by the European Research Training Network ``The Origin of
Planetary Systems'' (PLANETS, contract number HPRN-CT-2002-00308). 

\vspace{5cm}
{\it Facilities:} \facility{Spitzer (IRAC,MIPS,IRS)}, \facility{CAHA 2.2/CAFOS}, 
\facility{WHT (ISIS,WYFFOS)}, \facility{INT (IDS)},  \facility{Palomar/DBSP },
\facility{Pico Veleta/MAMBO}.


\begin{table}[htbp]
{\scriptsize
   \centering
   \caption{Cold disk candidates and {\it Spitzer} observing log} 
   \begin{tabular}{lllllclllll} 
      \hline \hline
   &                          &       &        &      & Prog- &    & \multicolumn{2}{l}{IRS} & \multicolumn{2}{l}{MIPS-70} \\
ID &   Name                   & Cloud & $\alpha_{K-24}$ & $\alpha_{8-24}$ & ram$^1$ & Refs.  & AOR           & Obs. Date    &  AOR  & Obs. Date  \\ \hline
1  & SSTc2d J032835.1+302010  & Per  &   0.15 & 2.13  & 1 & 1,15      & 19059712 & 2007-03-18 &  19954688  &  2007-02-03 \\
2  & ASR 118                                    & Per  &   -0.22 & 0.77 & 1 & 2,15      & 19056128 & 2006-09-19 &  19057664  &  2007-09-20 \\
3  & SSTc2d J032903.9+305630  & Per  &   -0.17 & 1.25 & 1 & 15         & 19954176 & 2007-03-19 &  19954688  &  2007-02-03 \\

4  & SSTc2d J032924.1+311958  & Per  & 0.36  & 1.57  & 1 & 1,15  & 19052544 & 2007-03-22 &  19057664  &  2007-09-20 \\
5  & SSTc2d J032929.3+311835  & Per  & -0.66 & 1.30 & 1 &  2,15  & 19053312 & 2007-03-09 &  19057664  &  2007-09-20 \\

6  & SSTc2d J033027.2+302830  & Per  &   0.00  & 2.37  & 1 & 1,15     & 19953920 & 2007-03-09 &  19057664  &  2007-09-20 \\
7  & SSTc2d J033035.5+311559  & Per  &   -0.19 & 2.04  & 1 & 15        & 19053056 & 2007-03-09 &  19954688  &  2007-02-03 \\
8  & SSTc2d J033038.2+303212  & Per  &   0.35  & 2.98  & 1 & 15        & 19953664 & 2007-03-19 &  19057664  &  2007-09-20 \\
9  & LZK 21                                        & Per  &   -0.66 & 0.06  & 1 & 3,15    & 19052032 & 2007-03-09 &  19954688  & 2007-02-23 \\
10 & SSTc2d J034227.1+314433 & Per  & -0.93 & 0.74   & 1 & 15        & 19053568 & 2007-03-18 &  19954432  & 2007-09-18 \\
11 & Cl* IC 348 LRL 190                & Per  & -1.00   & 0.05  & 1 & 4,6,7,15 & 19054848 & 2007-03-18 &  19954432  & 2007-09-18 \\

12 & SSTc2d J034434.8+315655 & Per  & -1.11 & 0.65 & 1 & 4,5,6,15 & 19053824 & 2007-03-15 &  19954432  & 2007-09-18 \\
13 & Cl* IC 348 LRL 265                & Per  &  -0.09 & 0.33  & 1 & 15 & 19052288 & 2006-09-18 &  19954432  & 2007-09-18 \\

14 & Cl* IC 348 LRL 110                & Per  &  -0.60 & 0.50  & 1 & 4,6,7,15  & 19052288 & 2006-09-18 &  19954432  & 2007-09-18 \\
15 & LkH$\alpha$ 329                    & Per  &  -1.02 & -0.24 & 1 & 8,15    & 19056384 & 2007-03-09 &  19954432  & 2007-09-18 \\
16 & Hn 24                                        & Cha &  -0.85 & -0.16  & 1 & 9,16   & 19054592 & 2007-03-11 &  19058944 & 2006-08-15 \\
17 & Sz 84                                         & Lup  & -2.34  & -0.49 & 2 & 10,19 & 05644288 & 2004-03-25 &  5697024   & 2005-03-07 \\
18 & SSTc2d J161029.6-392215 & Lup  & -1.07  & 0.01  & 1 & 17       & 19051008 & 2007-03-19 &  19055360 & 2006-08-18 \\
19 & RX J1615.3-3255                   & Oph  & -1.64 & -0.10  & 2 & 11,19 & 15916800 & 2005-09-09 &   5703424  & 2005-03-07 \\
20 & SSTc2d J162148.5-234027  & Oph  & -0.33 & 0.32  & 1 & 18       & 15920896 & 2005-09-09 &  \ldots         & \dots             \\
21 & SSTc2d J162221.1-230403  & Oph  & 0.93 &  1.43  & 1 & 18       & 15920896 & 2005-09-09 &  \ldots         & \dots              \\
22 & SSTc2d J162245.4-243124  & Oph  & -0.83 & 0.69  & 1 & 18       & 15920641 & 2006-03-15 &  \ldots         & \dots              \\
23 & SSTc2d J162332.9-225847  & Oph  & -1.07 & -0.37 & 1 & 18       & 15920641 & 2006-03-15 &  \ldots         & \dots              \\
24 & SSTc2d J162506.9-235050  & Oph  & -1.10 & 0.68  & 1 & 18       & 19059200 & 2007-03-19 &  19954944 & 2006-09-01 \\
25 & DoAr 21                                     & Oph  & -1.16 & -0.43 & 3 & 12,13,18 & 12699392 & 2006-03-15 &  \ldots         & \dots         \\
26 & SSTc2d J182829.1+002756  & Ser  & -0.67 & -0.11 & 4 & 20       & 17885184 & 2007-04-24 &  \ldots         & \dots                \\
27 & SSTc2d J182858.1+001724  & Ser  & -1.92 & -2.97 & 4 & 20       & 17888768 & 2007-05-05 &  \ldots         & \dots                \\
28 & SSTc2d J182907.0+003838  & Ser  & -0.75 & -1.48 & 4 & 20       & 17884672 & 2007-04-28 &  \ldots         & \dots                \\
29 & SSTc2d J182911.5+002039  & Ser  & -0.85 & 0.03  & 4 & 20       & 17887744 & 2007-04-24 &  \ldots        & \dots                \\
30 & SSTc2d J182915.6+003912  & Ser  & -0.31 & -1.21 & 1 & 20       & 17884672 & 2007-04-28 & 19057152 & 2007-05-18  \\
31 & SSTc2d J182915.6+003923 & Ser  & -0.47 & -0.73  & 1 & 20       & 17885440 & 2007-04-24 &   \ldots         & \dots                \\
32 & SSTc2d J182935.6+003504  & Ser  & -1.02 & -0.40 & 1 & 20       & 17884160 & 2007-04-19 &  19057152 & 2007-05-18  \\
33 & SSTc2d J182936.2+004217  & Ser  & -0.68 & -0.32 & 4 & 20       & 17884160 & 2007-04-19 &   \ldots         & \dots                \\
34 & SSTc2d J182944.1+003356  & Ser  & -1.41 & -1.05 & 4 & 20       & 17886720 & 2007-04-24 &   \ldots         & \dots                \\
35 & SSTc2d J182947.3+003223  & Ser  & -0.90 & -1.50 & 1 & 20       & 17887488 & 2007-04-25 & 19057152 & 2007-05-18  \\
    \hline     
     \end{tabular}      
  \label{tab:sample} 
\tablerefs{1) \cite{Walawender2005}, 2) \cite{Getman2002}, 3) \cite{Liu1980}, 4) \cite{Luhman1998}, 5) \cite{Muench2007}, 
6) \cite{Preibisch2003}, 7) \cite{Preibisch2001}, 8) \cite{Herbig1988}, 9) \cite{Hartigan1993}, 
10) \cite{Schwartz1977}, 11) \cite{Krautter1997}, 12) \cite{Dolizde1959}, 13) \cite{Gagne2004}, 14) \cite{Aspin1994},
15) Lai et al. (in prep.), 16) \cite{Alcala2008}, 17) \cite{Merin2008}, 18) \cite{Allers2006}, 19) \cite{Padgett2006},
20) \cite{Harvey2007b}, 21) \cite{Brown2007}, 22) \cite{Brown2008}, 23) \cite{Kessler-Silacci2006}}
\tablenotetext{1}{Data taken in program: 1) {\it Spitzer} GO3 \#30843 (PI: Mer\'{i}n), 2) c2d WTTS survey \citep{Padgett2006}, 
3) {\it Spitzer} GTO \# 2 (PI: Houch), 4) {\it Spitzer} GO3 \#30223 (PI: Pontoppidan), 
5) c2d CTTS IRS survey \citep{Kessler-Silacci2006}}
}
\end{table}         

\begin{deluxetable}{lllllllllllllllll}
\rotate{}
\tabletypesize{\scriptsize}
\tablewidth{0pt}
\tablecaption{IRAC, MIPS and MAMBO 1.3 mm fluxes for the cold disk candidate sample \label{tab:phottable2}} 
\tablecolumns{9}
\tablehead{\colhead{No.} & 
\colhead{IRAC $3.6$} & 
\colhead{IRAC $4.5$} &
\colhead{IRAC $5.8$} & 
\colhead{IRAC $8.0$} &
\colhead{MIPS $24.0$} & 
\colhead{MIPS $70.0$} & 
\colhead{$1.3$ mm} \\
\colhead{} & 
\colhead{(mJy)} & 
\colhead{(mJy)} &
\colhead{(mJy)} & 
\colhead{(mJy)} &
\colhead{(mJy)} & 
\colhead{(mJy)} & 
\colhead{(mJy)}  }
\startdata
1  &      1.45e+00$\pm$1.52e-02&      1.18e+00$\pm$1.52e-02&      9.55e-01$\pm$2.08e-02&      1.21e+00$\pm$2.19e-02&      4.25e+01$\pm$2.44e-01&      1.22e+02$\pm$4.69e+00&    $<$1.00e+00\\
2  &      1.68e+01$\pm$2.74e-01&      1.47e+01$\pm$2.74e-01&      1.21e+01$\pm$1.02e-01&      1.75e+01$\pm$1.20e-01&      1.78e+02$\pm$9.73e-01&      3.40e+02$\pm$9.35e+01&      3.70e+00$\pm$9.00e-01\\
3  &      9.64e-01$\pm$7.40e-03&      9.87e-01$\pm$7.40e-03&      8.28e-01$\pm$1.15e-02&      8.94e-01$\pm$1.34e-02&      1.14e+01$\pm$1.21e-01&      4.00e+01$\pm$4.06e+00&     $\ldots$\\
 4  &      2.01e+00$\pm$3.51e-02&      2.00e+00$\pm$3.51e-02&      2.05e+00$\pm$3.10e-02&      3.28e+00$\pm$3.21e-02&      6.78e+01$\pm$4.41e-01&    $<$3.75e+01&      1.20e+00$\pm$1.30e+00\\
5  &      1.28e+01$\pm$2.68e-01&      9.33e+00$\pm$2.68e-01&      6.96e+00$\pm$6.71e-02&      6.19e+00$\pm$4.73e-02&      9.55e+01$\pm$6.05e-01&      3.06e+02$\pm$1.31e+01&      6.30e+00$\pm$1.10e+00\\
6  &      1.51e+01$\pm$1.43e-01&      1.50e+01$\pm$1.43e-01&      1.05e+01$\pm$6.43e-02&      8.42e+00$\pm$5.92e-02&      4.31e+02$\pm$2.33e+00&      1.22e+03$\pm$8.20e+00&    $<$7.00e-01\\
7  &      1.46e+00$\pm$7.30e-02&      1.18e+00$\pm$5.86e-02&      1.02e+00$\pm$6.46e-02&      1.38e+00$\pm$7.64e-02&      4.17e+01$\pm$3.86e+00&      1.45e+02$\pm$4.06e+00&      1.30e+00$\pm$5.00e-01\\
8  &      1.05e+00$\pm$1.54e-02&      1.03e+00$\pm$1.54e-02&      8.18e-01$\pm$1.96e-02&      6.43e-01$\pm$1.81e-02&      5.75e+01$\pm$3.29e-01&      2.88e+02$\pm$6.25e+00&     $\ldots$\\
9  &      9.37e+01$\pm$1.11e+00&      8.70e+01$\pm$1.11e+00&      7.81e+01$\pm$4.61e-01&      8.57e+01$\pm$7.27e-01&      3.07e+02$\pm$1.84e+00&      4.02e+02$\pm$8.60e+00&      1.30e+01$\pm$1.30e+00\\
10 &      1.32e+01$\pm$1.21e-01&      9.87e+00$\pm$1.21e-01&      7.23e+00$\pm$5.75e-02&      5.50e+00$\pm$4.83e-02&      4.68e+01$\pm$2.81e-01&    $<$8.39e+01&    $<$1.40e+00\\
11 &      6.68e+00$\pm$4.26e-02&      5.22e+00$\pm$4.26e-02&      4.21e+00$\pm$2.74e-02&      3.62e+00$\pm$4.19e-02&      1.39e+01$\pm$1.62e-01&      2.50e+01$\pm$6.60e+01&     $\ldots$\\
12 &      1.13e+01$\pm$6.18e-02&      8.65e+00$\pm$6.18e-02&      6.47e+00$\pm$3.54e-02&      4.56e+00$\pm$3.94e-02&      2.10e+01$\pm$1.53e-01&    $<$7.60e+01&    $<$1.40e+00\\
13 &      1.23e+01$\pm$1.71e-01&      1.43e+01$\pm$1.71e-01&      1.31e+01$\pm$1.14e-01&      1.41e+01$\pm$1.40e-01&      1.23e+02$\pm$8.81e-01&    $<$1.37e+02&      1.16e+01$\pm$1.00e+00\\
14 &      1.32e+01$\pm$2.18e-01&      1.15e+01$\pm$2.18e-01&      8.93e+00$\pm$1.04e-01&      1.15e+01$\pm$2.06e-01&      6.71e+01$\pm$5.85e-01&     $<$2.16e+03&    $<$1.46e+00\\
15 &      1.27e+02$\pm$1.62e+00&      1.02e+02$\pm$1.62e+00&      7.66e+01$\pm$4.10e-01&      7.07e+01$\pm$5.18e-01&      1.75e+02$\pm$8.47e-01&      1.31e+02$\pm$7.86e+00&    $<$1.20e+00\\
16 &      1.16e+02$\pm$6.70e+00&      8.65e+01$\pm$4.69e+00&      7.71e+01$\pm$3.99e+00&      9.68e+01$\pm$4.78e+00&      2.48e+02$\pm$2.30e+01&      3.18e+02$\pm$3.37e+00&     $\ldots$\\
17 &      4.20e+01$\pm$3.51e-02&      2.90e+01$\pm$3.51e-02&      2.01e+01$\pm$3.10e-02&      1.20e+01$\pm$3.21e-02&      2.09e+01$\pm$4.41e-01&      2.44e+02$\pm$ 7.32e+01 &    $<$3.60e+01\\
18 &      1.91e+01$\pm$9.30e-01&      1.42e+01$\pm$9.30e-01&      1.15e+01$\pm$5.60e-01&      1.09e+01$\pm$5.30e-01&      3.37e+01$\pm$3.18e+00&      1.10e+02$\pm$1.80e+01&     $\ldots$\\
19 &      1.14e+02$\pm$9.30e-01&      8.50e+01$\pm$9.30e-01&      6.10e+01$\pm$5.60e-01&      6.60e+01$\pm$5.30e-01&      2.71e+02$\pm$3.18e+00&      7.27e+02$\pm$1.80e+01&     $\ldots$\\
20 &      1.24e+01$\pm$6.00e-01&      1.20e+01$\pm$5.70e-01&      1.08e+01$\pm$5.20e-01&      1.52e+01$\pm$7.30e-01&      7.99e+01$\pm$7.39e+00&    $<$6.82e+01&    $<$4.60e+00\\
21 &      1.61e+00$\pm$8.00e-02&      1.60e+00$\pm$7.00e-02&      1.78e+00$\pm$1.00e-01&      7.72e+00$\pm$3.80e-01&      1.17e+02$\pm$1.08e+01&    $<$5.38e+02&    $<$4.40e+00\\
22 &      9.21e+01$\pm$4.58e+00&      6.15e+01$\pm$3.04e+00&      4.47e+01$\pm$2.14e+00&      5.13e+01$\pm$2.47e+00&      3.45e+02$\pm$3.20e+01&    $<$8.31e+01&    $<$5.40e+00\\
23 &      3.23e+01$\pm$1.61e+00&      2.44e+01$\pm$1.17e+00&      1.92e+01$\pm$9.20e-01&      2.23e+01$\pm$1.07e+00&      4.79e+01$\pm$4.47e+00&    $<$4.32e+02&     $\ldots$\\
24 &      6.07e+01$\pm$2.74e-01&      4.17e+01$\pm$2.74e-01&      3.09e+01$\pm$1.02e-01&      2.28e+01$\pm$1.20e-01&      1.59e+02$\pm$9.73e-01&      5.37e+02$\pm$7.88e+01&     $\ldots$\\
25 &      1.19e+03$\pm$1.94e+01&      8.45e+02$\pm$1.49e+01&      7.30e+02$\pm$1.00e+01&      6.39e+02$\pm$8.50e+00&      1.34e+03$\pm$3.42e+01&      1.20e+04$\pm$1.26e+03&     $\ldots$\\
26 &      9.80e+00$\pm$5.75e-01&      8.51e+00$\pm$5.75e-01&      7.03e+00$\pm$3.82e-01&      5.46e+00$\pm$2.98e-01&      4.34e+01$\pm$4.03e+00&    $<$2.87e+02$^*$&      2.40e+01$\pm$6.00e-01\\
27 &      5.25e+01$\pm$2.97e+00&      3.67e+01$\pm$2.97e+00&      3.12e+01$\pm$1.68e+00&      2.85e+01$\pm$1.68e+00&      9.74e+00$\pm$9.20e-01&      1.17e+03$\pm$1.01e+02&     $\ldots$\\
28 &      2.13e+01$\pm$1.43e+00&      2.93e+01$\pm$1.43e+00&      2.25e+01$\pm$1.29e+00&      2.27e+01$\pm$1.10e+00&      4.02e+01$\pm$3.72e+00&    $<$2.87e+02$^*$&     $\ldots$\\
29 &      4.46e+00$\pm$2.73e-01&      3.36e+00$\pm$2.73e-01&      2.39e+00$\pm$1.47e-01&      1.72e+00$\pm$1.03e-01&      1.60e+01$\pm$1.49e+00&    $<$2.87e+02$^*$&     $\ldots$\\
30 &      1.36e+01$\pm$8.67e-01&      1.62e+01$\pm$8.67e-01&      1.51e+01$\pm$8.86e-01&      2.56e+01$\pm$1.26e+00&      6.05e+01$\pm$5.59e+00&    $<$5.00e+01&     $\ldots$\\
31 &      6.42e+00$\pm$4.61e-01&      7.33e+00$\pm$4.61e-01&      5.99e+00$\pm$3.46e-01&      8.05e+00$\pm$3.94e-01&      3.25e+01$\pm$3.01e+00&      1.35e+02$\pm$1.57e+01&     $\ldots$\\
32 &      3.05e+01$\pm$2.00e+00&      2.41e+01$\pm$2.00e+00&      1.70e+01$\pm$9.68e-01&      1.28e+01$\pm$6.61e-01&      7.42e+01$\pm$6.86e+00&      5.76e+01$\pm$1.31e+01&     $\ldots$\\
33 &      2.11e+01$\pm$1.24e+00&      1.99e+01$\pm$1.24e+00&      1.56e+01$\pm$8.31e-01&      1.28e+01$\pm$6.19e-01&      8.11e+01$\pm$7.49e+00&      1.26e+02$\pm$1.74e+01&      2.30e+00$\pm$6.00e-01\\
34 &      2.08e+01$\pm$1.19e+00&      1.37e+01$\pm$1.19e+00&      1.10e+01$\pm$5.91e-01&      6.97e+00$\pm$3.45e-01&      1.97e+01$\pm$1.83e+00&      1.40e+02$\pm$1.71e+01&      6.30e+00$\pm$6.00e-01\\
35 &      1.04e+01$\pm$6.24e-01&      9.28e+00$\pm$6.24e-01&      8.20e+00$\pm$4.53e-01&      1.06e+01$\pm$5.43e-01&      1.82e+01$\pm$1.69e+00&      1.76e+01$\pm$6.81e+00&      2.20e+00$\pm$9.00e-01\\
\enddata
\tablenotetext{*}{Upper limits are estimates from nearby sources due to difficulties with the local background flux.}
\end{deluxetable}

\setlength{\voffset}{30mm}
\begin{deluxetable}{lccccl}
\tablecaption{Optical spectroscopy observation log}     
\tablecolumns{6}
\tabletypesize{\footnotesize}
\tablehead{\colhead{Date} & \colhead{Telescope +}       & \colhead{Grism}  & \colhead{Wavelength}  & \colhead{Dispersion}  & colhead{ID}  \\
                               &  \colhead{Instrument}        &           &  \colhead{Coverage}     &  \colhead{(\AA/resel)} & }
\startdata
2006 November 22  &  WHT+WYFFOS  &  R600R   &  5500--8500     &  4   &   2, 3, (4), 5, 7, (10), (11)      \\
                                &   &             &      &           &   12, 13, 14, 15      \\
\hline
2006 December 13   &                          &                     &      &                      &    1, (9)       \\
2007 July 26              &      WHT+ISIS    &   R600R     &     6000--8000      &          &     24      \\
2008 July 22              &                          &                           &    &                        &      2      \\
\hline
2007 May 06              &        INT+IDS     & R600R          &    4700--8700    &          &     22, 24      \\
\hline
2008 June 23-25       &      2.2m/CAFOS       &  R100     &      6000--9000      &  2  &    21, 23, 24, 25, 29        \\
\hline
2008  December 30   &      Palomar/DBSP      & R316      &      6200--8700       & 10 &    4, 6, 8      \\
2009    July 21           &      Palomar/DBSP      & R316      &      6200--8700       &      &   9, 10     \\
\hline
2009 October 30       &      WHT+ISIS     & R316              &     5500--8300       &  & 10, 13    \\
\hline
2009 August 03      &      CTIO 1.5m/R-C Spec     & 47/Ib    &     5630--6930       &  & 18,19   \\
\enddata
\tablenotetext{1}{()=low S/N}
\tablenotetext{2}{fiber size}

\label{t_obs}

\end{deluxetable}

\begin{deluxetable}{l l c c c c l c c c c c c c c}
  \tabletypesize{\tiny} \tablewidth{0pt} \tablecaption{Stellar parameters for the stars in the    sample.\label{tab:stellarparameters}} 
  \tablecolumns{13}
  \tablehead{\colhead{No.} & \colhead{Object Id.} &\colhead{SpT} &
    \colhead{Ref} & \colhead{A$_V$} & \colhead{T$_{\rm eff}$} &
    \colhead{L$_*$} & \colhead{Mass} &
    \colhead{EW[H$\alpha$]$^*$}
    &\colhead{H$\alpha$[10\%]} & \colhead{$\log\dot{M_{acc}}$} \\
    \colhead{} & \colhead{} & \colhead{} & \colhead{} &
    \colhead{(mag)} & \colhead{(K)} & \colhead{(L$_\odot$)} & &
    \colhead{(\AA)} & \colhead{(km/s)} &
    \colhead{(M$_\odot$/yr)} & } 
    \startdata 
    2 & ASR 118 & K4 & 1 & 13.0 & 4590  & 1.17   & 1.36   & 26         & 293    & -10.2  \\
  5 & SSTc2d J032929.3+311835 & M0 & 1 & \ldots & 3850  & 0.09 & \ldots & 4.8        & \ldots & \ldots  \\
  10 & SSTc2d J034227.1+314433 & K7 & 1 & 5.0 & 4060  & 0.22   & 0.82   & 4.3        & \ldots & \ldots \\
  11 & Cl* IC 348 LRL 190 & M3.75 & 6 & 6.5 & 3306  & 0.09   & 0.35   & 5          & \ldots & \ldots \\
  12 & SSTc2d J034434.8+315655 & M3 & 1 & 3.0 & 3415  & 0.12   & 0.45   & 130        & 504    & -8.0   \\
  17 & Sz 84 & M5.5 & 3 & 0.5 & 3057  & 0.15   & 0.14   & 44         & \ldots & \ldots \\
  18 & SSTc2d J161029.6-392215 & M5 & 1 & 1.0 & 3125  & 0.08   & 0.21   & 18         & \ldots & \ldots \\
  19 & RX J1615.3-3255 & K4 & 1 & 1.0 & 4590  & 0.85   & 1.28   & 26         & \ldots & \ldots \\
  22 & SSTc2d J162245.4-243124 & M2 & 1 & 2.0 & 3580  & 0.29   & 0.59   & 5.0        & 224    & \ldots \\
  24 & SSTc2d J162506.9-235050 & M0.5 & 1 & 3.0 & 3778  & 0.21   & 0.77   & 4.6        & 505    & -8.0   \\
  26 & SSTc2d J182829.1+002756 & K7 & 2 & 13.0 & 4060  & 0.34   & 0.94   & \ldots     & \ldots & \ldots \\
  29 & SSTc2d J182911.5+002039 & M0 & 13 & 6.0 & 3850  & 0.13   & 0.72   & 25         & 656    & -6.5   \\
  32 & SSTc2d J182935.6+003504 & K7 & 13 & 4.0 & 4060  & 0.52   & 1.06   & 10.9       & 273    & -10.2  \\
  33 & SSTc2d J182936.2+004217 & F9 & 13 & 11.0 & 6115  & 1.89   & 1.30   & \ldots     & \ldots & \ldots \\
  34 & SSTc2d J182944.1+003356 & M0 & 13 & 5.0 & 3850  & 0.54   & 0.93   & 37         & 444    & -8.6   \\  \hline 
  1 & SSTc2d J032835.1+302010$^\dagger$ & K4 & 1 & 0.0 & 4590 & 0.07$^*$  & \ldots & 8.1        & 449    & -8.5   \\
  3 & SSTc2d J032903.9+305630$^\dagger$ & G7 & 1 & 6.0 & 5630  & 0.05$^*$   & \ldots    & 54         & 344    & -9.6   \\
  4 & SSTc2d J032924.1+311958$^\dagger$ & K5 & 1 & 5.0 & 4350  & 0.41   & 0.96   & 6.5        & 262    & -10.4 \\
  6 & SSTc2d J033027.2+302830$^{\dagger,a}$ & c$^b$ & 1 & \ldots &  \ldots & \ldots & \ldots &  106        & \ldots & \ldots \\
  7 & SSTc2d J033035.5+311559$^\dagger$ & K5 & 1 & \ldots & 4350  & 0.07$^*$ & \ldots & 15.1       & 405    & -9.0   \\
  8 & SSTc2d J033038.2+303212$^\dagger$ & M0$^c$ & 1 & \ldots & 3850  & 0.05$^*$  & \ldots &  \ldots     & 194    & \ldots \\
  9 & LZK 21 & M0 & 1 & 6.0 & 3850  & 1.46   & 1.00   & 72         & \ldots & \ldots \\
 13 & Cl* IC 348 LRL 265 & M3 & 1 & 17.  & 3415  & 0.42       &  0.52      & 113        & \ldots & \ldots \\
  14 & Cl* IC 348 LRL 110 & M0.5 & 1 & 5.0 & 3778  & 0.22   & 0.78   &  36         & 424    & -8.8   \\
  15 & LkH$\alpha$ 329 & K5 & 1 & 5.0 & 4350  & 2.89   & 1.63   &19.4       & 700    & -6.1   \\
  16 & Hn 24 & M0.5 & 1 & 2.0 & 3778  & 0.75   & 0.92   & 0.2        & 311    & \ldots \\
  20 & SSTc2d J162148.5-234027 & M5.5 & 5 & 7.0 & 3057  & 0.04   & 0.14  & \ldots$^d$ & \ldots & \ldots \\
  21 & SSTc2d J162221.1-230403$^\dagger$ & K5 & 1 & \dots & 4350  & 0.02$^*$  & \dots  & 45.8       & 594    & -7.1   \\
  23 & SSTc2d J162332.9-225847 & M0.5 & 1 & 5.0 & 3778  & 0.15   & 0.73  & 2.8        & 413    & \ldots \\
  25 & DoAr 21 & K1 & 4 & 6.0 & 5080  & 9.84   & 1.80   & 1--2       & 450    & \ldots \\
  27 & SSTc2d J182858.1+001724 & G3 & 13 & 6.0 & 5830  & 4.76   & 1.4    & -2.2       & \ldots & \ldots \\
  28 & SSTc2d J182907.0+003838 & K7 & 2 & 16.0 & 4060  & 0.6    & 1.04   & \ldots     & \ldots & \ldots \\
  30 & SSTc2d J182915.6+003912 & K5 & 13 & 12.0 & 4350  & 0.45   & 1.00   & 22         & 532    & -7.7   \\
  31 & SSTc2d J182915.6+003923$^\dagger$ & K7 & 2 & 0.0 & 4060  & 0.20   & \ldots  & \ldots     & \ldots & \ldots \\
  35 & SSTc2d J182947.3+003223 & M0 & 13 & 7.0 & 3850  & 0.28   & 0.85   & 49         & 618    & -6.9   \\
\enddata
\tablecomments{For clarity, cold disks are shown above the horizontal
  line while other objects are below.}  \tablecomments{We used the c2d
  distances to the clouds to compute the stellar luminosities: 250 pc
  for Perseus, 187 pc for Cham. II, 150 pc for Lupus II and III, 120
  pc for Ophiuchus and 260 pc for Serpens.}
\tablenotetext{*}{H$\alpha$ equivalent widths are listed as positive
  when seen in emission and negative when seen in absorption}
\tablenotetext{\dagger}{: Objects identified as likely edge-on
  disks. Luminosities are presented to show that the stars are
  underluminous. In these cases, we do not calculate masses and ages.}
\tablenotetext{a}{: Source is spatially extended in slit.}
\tablenotetext{b}{: Continuum spectral type. The object is heavily
  veiled and no SpT was obtainable at low spectral resolution.}
\tablenotetext{c}{: Large uncertainty due to poor S/N and large
  veiling.}  \tablenotetext{d}{: Accreting based on presence of
  Paschen $\beta$ emission (Gully-Santiago et al. in prep.).}  \tablerefs{1)
  Spectral type from this work with optical spectroscopy; 2) Spectral
  type from this work with photometry; 3) \cite{Krautter1997}; 4)
  \cite{Jensen2009} 5) Gully-Santiago et al. in prep. 6)
  \cite{Luhman2003} 13) \cite{Oliveira2009}}
\end{deluxetable}

\begin{deluxetable}{lllllllllllllllll}
\tabletypesize{\scriptsize}
\tablewidth{0pt}
\tablecaption{Optical and near-IR photometry for the cold disk candidate sample \label{tab:phottable1}} 
\tablecolumns{9}
\tablehead{\colhead{No.} & 
\colhead{$U$} & 
\colhead{$B$} &
\colhead{$V$} & 
\colhead{$R_C$} &
\colhead{$I_C$} & 
\colhead{$J_{2MASS}$} & 
\colhead{$H_{2MASS}$} &
\colhead{$K_{2MASS}$} \\
\colhead{} & 
\colhead{(mag)} & 
\colhead{(mag)} &
\colhead{(mag)} & 
\colhead{(mag)} &
\colhead{(mag)} & 
\colhead{(mag)} & 
\colhead{(mag)} &
\colhead{(mag)} }
\startdata
1  &  $\ldots$&  $\ldots$&  $\ldots$&  $\ldots$&  $\ldots$&   15.66$\pm$0.05&   14.41$\pm$0.05&   13.69$\pm$0.05\\
2  &  $\ldots$&  $\ldots$&  $\ldots$&  $\ldots$&  $\ldots$&   13.76$\pm$0.05&   11.92$\pm$0.05&   11.08$\pm$0.05\\
3  &  $\ldots$&  $\ldots$&  $\ldots$&  $\ldots$&  $\ldots$&   15.64$\pm$0.05&   14.64$\pm$0.05&   14.37$\pm$0.05\\
4  &  $\ldots$&  $\ldots$&  $\ldots$&  $\ldots$&  $\ldots$&   15.52$\pm$0.05&   14.25$\pm$0.05&   13.57$\pm$0.05\\
5  &  $\ldots$&  $\ldots$&  $\ldots$&  $\ldots$&  $\ldots$&   12.59$\pm$0.05&   11.37$\pm$0.05&   10.96$\pm$0.05\\
6  &  $\ldots$&  $\ldots$&  $\ldots$&  $\ldots$&  $\ldots$&   14.86$\pm$0.05&   13.16$\pm$0.05&   11.90$\pm$0.05\\
7  &  $\ldots$&  $\ldots$&  $\ldots$&  $\ldots$&  $\ldots$&   15.66$\pm$0.09&   14.40$\pm$0.06&   13.69$\pm$0.05\\
8  &  $\ldots$&  $\ldots$&  $\ldots$&  $\ldots$&  $\ldots$&   15.57$\pm$0.05&   14.65$\pm$0.05&   14.04$\pm$0.05\\
9  &  $\ldots$&  $\ldots$&  $\ldots$&  $\ldots$&  $\ldots$&   11.29$\pm$0.05&   10.05$\pm$0.05&    9.41$\pm$0.05\\
10 &  $\ldots$&  $\ldots$&  $\ldots$&  $\ldots$&  16.33$\pm$0.01 &   13.38$\pm$0.05&   11.86$\pm$0.05&   11.15$\pm$0.05\\
11 &  $\ldots$&  $\ldots$&  $\ldots$&  20.45$\pm$0.05 &  17.93$\pm$0.05 &   14.38$\pm$0.05&   12.87$\pm$0.05&   12.11$\pm$0.05\\
12 &  $\ldots$&  $\ldots$&  $\ldots$&  $\ldots$&  $\ldots$&   13.02$\pm$0.05&   12.09$\pm$0.05&   11.63$\pm$0.05\\
13 &  $\ldots$&  $\ldots$&  $\ldots$&  $\ldots$&  $\ldots$&   15.64$\pm$0.05&   13.37$\pm$0.05&   12.00$\pm$0.05\\
14 &  $\ldots$&  $\ldots$&  $\ldots$&  $\ldots$&  15.78$\pm$0.05 &   13.07$\pm$0.05&   11.85$\pm$0.05&   11.28$\pm$0.05\\
15 &  $\ldots$&  $\ldots$&  $\ldots$&  $\ldots$&  $\ldots$&   10.46$\pm$0.05&    9.46$\pm$0.05&    8.93$\pm$0.05\\
16 &  $\ldots$&  $\ldots$&  $\ldots$&  $\ldots$&  $\ldots$&   10.15$\pm$0.05&    9.33$\pm$0.05&    8.88$\pm$0.05\\
17 &  $\ldots$&   17.54$\pm$0.03&   16.06$\pm$0.11&   14.53$\pm$0.04&   12.94$\pm$0.02&   10.93$\pm$0.02&   10.20$\pm$0.02&    9.85$\pm$0.02\\
18 &  $\ldots$&   17.51$\pm$0.03&   16.31$\pm$0.01&  15.79$\pm$0.05 &  13.90$\pm$0.05 &   11.94$\pm$0.05&   11.27$\pm$0.05&   10.91$\pm$0.05\\
19 &  $\ldots$&  $\ldots$&   12.04$\pm$0.01&  11.28$\pm$0.05 &  10.54$\pm$0.05 &    9.55$\pm$0.05&    8.78$\pm$0.05&    8.56$\pm$0.05\\
20 &  $\ldots$&  $\ldots$&  $\ldots$&   19.59$\pm$0.04&  $\ldots$&   13.59$\pm$0.05&   12.35$\pm$0.05&   11.69$\pm$0.05\\
21 &  $\ldots$&   19.18$\pm$0.03&   17.74$\pm$0.01&   17.03$\pm$0.04&  $\ldots$&   14.66$\pm$0.05&   13.87$\pm$0.05&   13.57$\pm$0.05\\
22 &  $\ldots$&  $\ldots$&   15.74$\pm$0.50&  14.82$\pm$0.05 &  12.51$\pm$0.02 &   10.37$\pm$0.05&    9.43$\pm$0.05&    9.06$\pm$0.05\\
23 &  $\ldots$&  $\ldots$&  $\ldots$&   16.22$\pm$0.04&  $\ldots$&   11.49$\pm$0.05&   10.68$\pm$0.05&   10.29$\pm$0.05\\
24 &  $\ldots$&   18.38$\pm$0.05&   16.59$\pm$0.05&   15.55$\pm$0.05&  $\ldots$&   11.06$\pm$0.05&   10.03$\pm$0.05&    9.51$\pm$0.05\\
25 &  $\ldots$&   15.80$\pm$0.75&   14.00$\pm$0.50&   12.00$\pm$0.05&   10.05$\pm$0.90&    8.09$\pm$0.02&    6.86$\pm$0.05&    6.23$\pm$0.02\\
26 &  $\ldots$&  $\ldots$&  $\ldots$&  $\ldots$&  $\ldots$&   14.99$\pm$0.05&   12.76$\pm$0.05&   11.81$\pm$0.05\\
27 &  $\ldots$&  $\ldots$&  $\ldots$&  $\ldots$&  $\ldots$&   10.72$\pm$0.05&    9.83$\pm$0.05&    9.45$\pm$0.05\\
28 &  $\ldots$&  $\ldots$&  $\ldots$&  $\ldots$&  $\ldots$&   15.20$\pm$0.05&   12.90$\pm$0.05&   11.55$\pm$0.05\\
29 &  $\ldots$&  $\ldots$&  $\ldots$&  $\ldots$&  $\ldots$&   14.03$\pm$0.05&   12.89$\pm$0.05&   12.42$\pm$0.05\\
30 &  $\ldots$&  $\ldots$&  $\ldots$&  $\ldots$&  $\ldots$&   14.53$\pm$0.05&   12.65$\pm$0.05&   11.69$\pm$0.05\\
31 &  $\ldots$&  $\ldots$&  $\ldots$&   18.38$\pm$0.05&  $\ldots$&   14.14$\pm$0.05&   12.79$\pm$0.05&   12.13$\pm$0.05\\
32 &  $\ldots$&  $\ldots$&  $\ldots$&  $\ldots$&  $\ldots$&   12.00$\pm$0.05&   10.84$\pm$0.05&   10.26$\pm$0.05\\
33 &  $\ldots$&  $\ldots$&  $\ldots$&  $\ldots$&  $\ldots$&   13.27$\pm$0.05&   11.92$\pm$0.05&   11.31$\pm$0.05\\
34 &  $\ldots$&  $\ldots$&  $\ldots$&   19.04$\pm$0.50&  $\ldots$&   12.19$\pm$0.05&   11.15$\pm$0.05&   10.77$\pm$0.05\\
35 &  $\ldots$&  $\ldots$&  $\ldots$&   19.14$\pm$0.05&  $\ldots$&   13.45$\pm$0.05&   12.11$\pm$0.05&   11.45$\pm$0.05\\
\enddata

\tablerefs{Optical fluxes come from 1) Bouy et al. (in prep) for Perseus sources (1-15); 2) \cite{Spezzi2008} for the only object from Chamaeleon II (\#16); 3) \cite{Comeron2009} for the Lupus sources (\# 17 and 18); 4) Bouy et al. (in prep.) for Ophiuchus sources (19 - 25); and 5) \cite{Spezzi2010} for the sources in Serpens (26-35)}
\end{deluxetable}

\begin{deluxetable}{l c c c c c}
\tabletypesize{\tiny}
\tablewidth{0pt}
\tablecaption{Selection criteria for edge-on disks\label{tab:edgeon}}
\tablecolumns{6}
\tablehead{\colhead{No.} & \colhead{Under} &\colhead{Forbidden} &
  \colhead{Silicate} & \colhead{Ice} & \colhead{High}\\
\colhead{} & \colhead{luminous} &\colhead{lines} &
  \colhead{absorption} & \colhead{absorption} & \colhead{L$_{\rm Disk}$/L$_*$}}
\startdata
 1  & \checkmark &      &  &  & \checkmark \\
 3  & \checkmark & \checkmark ? & \checkmark & \checkmark & \\
 4  &  &  & \checkmark  & \checkmark & \\
 6  &  & \checkmark &  &  & \checkmark\\
 7  & \checkmark &  & \checkmark  &  & \checkmark\\
 8  & \checkmark & \checkmark  & \checkmark  &  & \checkmark \\
 21 & \checkmark & \checkmark &  &  & \checkmark \\
 31 &  &  & & & \checkmark ? \\
\enddata
\end{deluxetable}

\begin{deluxetable}{r c l c c c c c c c c c c }
\tabletypesize{\scriptsize}
\tablewidth{0pt}
\tablecaption{Disk physical parameters and observables \label{tab:diskparams}} 
\tablecolumns{13}
\tablehead{\colhead{No.} & \colhead{Reg.} & \colhead{R$_{\rm hole}$$^{*}$} & \colhead{Inc.} & \colhead{Mass$^\dagger$} & \colhead{Sett.} & 
\colhead{F$_{mm}$} & 
\colhead{M$_{\rm disk}$} & 
\colhead{Ref.} &
\colhead{F$_{30}$/F$_{13}$} & 
\colhead{L$_{\rm disk}$/L$_*$} &
\colhead{R$_{\rm hole}^{\rm Rob.}$} &
\colhead{Model ID}  \\  
\colhead{} & 
\colhead{} & 
 \colhead{(AU)} &
   \colhead{(deg)} & 
      \colhead{10$^{-3}$ M$_\odot$} & 
      \colhead{} & 
   \colhead{(mJy)} &
   \colhead{10$^{-3}$ M$_\odot$} & 
\colhead{} &
\colhead{} & 
\colhead{} & 
\colhead{(AU)} &
\colhead{(Rob.)}}
\startdata
  1 & $\ldots$ & edge & \ldots & \ldots & \ldots & $<$1.0         & $<$0.3              & 1 &  20.17 & \ldots & \ldots & \ldots \\
  2 & B & 1.5$\pm$1 & 75 & obs & 0.5 & 3.7$\pm$0.9 & 1.1$\pm$0.3      & 1 & 4.29 & 0.16 & 1.64 & 3005228 \\
  3 & B & edge & \ldots & \ldots & \ldots & \ldots         & \ldots        & \ldots        & 21.35 & \ldots & \ldots &  \ldots \\
  4 & $\ldots$ & edge & \ldots & \ldots & \ldots & $<$1.3 & $<$0.4   & 1 &  9.82 & \ldots & \ldots &  \ldots \\
  5 & A & 6$^{\rm+4}_{\rm-2}$ & 10 & obs & 0.5 & 6.3$\pm$1.1 & 1.9$\pm$0.3    & 1 & 12.86 & 0.2 & 15.3 & 3000829 \\
  6 & $\ldots$ & edge & \ldots & \ldots & \ldots & $<$0.7         & $<$0.2            & 1 &   17.63 & \ldots & \ldots &  \ldots  \\
  7 & B & edge & \ldots & \ldots & \ldots &1.3$\pm$0.5 & 0.4$\pm$0.2 & 1 &  17.47 & \ldots & \ldots &  \ldots \\
  8 &$\ldots$ &  edge & \ldots & \ldots & \ldots & \ldots & \ldots      & \ldots &  33.19 & \ldots & \ldots &  \ldots  \\
  9 & B & no hole & 30 & obs & 0.5 & 13.0$\pm$1.3 & 4.0$\pm$0.4 & 1&  3.01 & 0.19 & \ldots & \ldots \\
 10 & A & 5$\pm$1 & 60 & 0.001 & 1 & $<$1.4  & $<$0.4                   & 1 & 4.76 & 0.04 & 1.87 & 3018041 \\
 11 & A & 5$\pm$1 & 30 & 0.003 & 1 &  \ldots       & \ldots               & \ldots &  7.85 & 0.22 &0.75 & 3015642 \\
 13 & $\ldots$ & no hole & 45 & obs & 1 & 11.6$\pm$1.0 & 3.5$\pm$0.3  & 1 &  4.51 & 0.18 &  1.71 & 3007226 \\
 12 & A & 3$\pm$1 &  5 & 0.002 & 1 & $<$1.4    & $<$0.4                   & 1  & 5.9 & 0.2 & 1.04 & 3017747 \\
 14 & B & no hole & 60 & 0.3 & 1 & $<$1.5  & $<$0.5                   & 1 &  5.78 & 0.23 & 0.8 & 3019220 \\
 15 & B & no hole & 75 & 0.005 & 1 & $<$1.2  & $<$0.4                   & 1     &  2.38 & 0.05 &  8.6 & 3005263 \\
 16 & B & no hole & 65 & 0.05 & 1 & \ldots & \ldots                     & \ldots &  2.65 & 0.12 &  $\ldots$ & 3012969 \\
 17 & $\ldots$ & 55$\pm$5 & 45 & 0.03 & 1 & $<$35.9    & $<$4.0                & 2 &  11.47 & 0.15 & 96.0 & 3004704 \\
 18 & A & 2$\pm$1 & 75 & 0.5 & 0.05 & \ldots & \ldots                     & \ldots  &  4.67 & 0.09 & $\ldots$ & 3000691 \\
 19 & A & 2$\pm$1 & 5 & 0.5 & 0.1 & \ldots & \ldots                     & \ldots &  5.36 & 0.09 & 3.8 & 3017818 \\
 20 & $\ldots$ & no hole & 5 & 0.01 & 1 & $<$4.6  & $<$0.3                     & 1 &  3.82 & 0.33 & 0.8 & 3019048 \\
 21 & $\ldots$ & edge & \ldots & \ldots & \ldots & $<$4.4 & $<$0.3                   & 1 &  5.55 & \ldots & \ldots & \ldots \\
 22 & A & 1$^{\rm+1}_{\rm-0.5}$ & 30 & 0.25 & 1 & $<$5.4  & $<$0.4                  & 1 &  3.16 & 0.18 & 2.75 & 3010142 \\
 23 & B & no hole & 80 & 0.004 & 1 & \ldots & \ldots                & \ldots &  3.55 & 0.19 &  0.3 & 3004623 \\
 24 & A & 3$\pm$2  & 5 & 0.1 & 0.25  & \ldots & \ldots                & \ldots &  6.98 & 0.13 & 2.21 & 3005973 \\
 25 & A & back  & \ldots &  \ldots & \ldots & \ldots & \ldots               & \ldots & 5.67 & 0.12 & \ldots & \ldots  \\
 26 & A & 4$\pm$1 & 5 & obs & 0.15 & 24.0$\pm$0.6 & 8.0$\pm$0.2 & 1  & 8.87 & 0.17 & 17.6 & 3002156 \\
 27 & $\ldots$ & back & \ldots & \ldots & \ldots &  \ldots & \ldots    & \ldots & 1.28 & 0.03 & \ldots & \ldots \\
 28 & $\ldots$ & no hole & 5 & 0.002 & 1 &  \ldots & \ldots                    & \ldots  &  2.41 & 0.15 & $\ldots$ & 3008873 \\
 29 & A & 8$\pm$2 & 30 & 0.01 & 1 & \ldots & \ldots                 & \ldots  &   11.71 & 0.07 & 3.76 & 3010258 \\
 30 & $\ldots$ & no hole & 45 & 0.05 & 1     & \ldots & \ldots                  & \ldots  &  2.53 & 0.03 & 0.5 & 3006320  \\
 31 & $\ldots$ & edge & 81.4 & \ldots & \ldots & \ldots & \ldots                 & \ldots  &  5.89 & \ldots & \ldots & \ldots \\
 32 & A & 7$\pm$1 & 5 & 0.01 & 1 & \ldots & \ldots                      & \ldots  &  6.01 & 0.11 & 4.14 & 3008745 \\
 33 & A & 13$\pm$3 & 5 & obs & 0.005 & 2.3$\pm$0.6 & 0.8$\pm$0.2    & 1 &  7.36 & 0.33 & 3.94 & 3018536 \\
 34 & A & 25$^{\rm+15}_{\rm-5}$ & 5 & obs & 0.25 & 6.3$\pm$0.6 & 2.1$\pm$0.2 & 1 &  10.63 & 0.02 & 8.94 & 3019329 \\
 35 & $\ldots$ & no hole & 80 & obs & 0.025 & 2.2$\pm$0.9 & 0.7$\pm$0.3 & 1 &  1.5 & 0.13 &  0.4 & 3016446 \\
\enddata

\tablerefs{1) Continuum flux at 1.3 mm from this work (\S~\ref{mmobs}); 2) \cite{Nuernberger1997};  3) \cite{Osterloh1995}; 4) \cite{Andre1990}; 5) \cite{Henning1993}; 6) \cite{Sylvester1996}.}
\tablecomments{Column "Ref." gives the regions in the upper left panel of Figure \ref{fig:criteria}.}
\tablecomments{$^*$: ``edge'' labels edge-on disks; ``back'' labels the two SEDs contaminated by background cloud material.}
\tablecomments{$^\dagger$: obs indicates that the disk mass in the model is the observed millimeter mass.}
\end{deluxetable}

\begin{table}
{\scriptsize
 \caption{Classification and mineralogy of the disk sample  \label{tab:sampleclass}} 
\begin{tabular}{l l c c c c c c c }
\hline\hline
  \multicolumn{1}{l}{No.} &
  \multicolumn{1}{l}{Name} &
  \multicolumn{1}{c}{Hole} &
  \multicolumn{1}{c}{Acc.} &
  \multicolumn{1}{c}{M$_{\rm disk}$} &
  \multicolumn{1}{c}{10 $\mu$m} &
  \multicolumn{1}{c}{20 $\mu$m} &
  \multicolumn{1}{c}{PAH} &
  \multicolumn{1}{c}{Cryst.} \\
\hline
  2 & ASR118                                                      & Y & Y & L & Y & Y & N & T\\
 5 & SSTc2d J032929.3+3118                     & Y & N & L & Y & Y & N & T\\
 10 & SSTc2d J034227.1+3144                    & Y & N & L & Y & N & N & Y\\
  11 & Cl* IC 348 LRL 190                               & Y & N & \ldots & Y & Y & Y & N\\
  12 & SSTc2d J034434.8+3156                    & Y & Y & L & Y & Y & N & Y\\
  17 & Sz 84                                                        & Y$^1$ & Y & L & N & N & N & N\\
  18 & SSTc2d J161029.6-3922                     & Y & \ldots & \ldots & Y & Y & N & Y\\
  19 & RX J1615.3-3255                                  & Y & Y & \ldots & Y & Y & Y & N\\
  22 & SSTc2d J162245.4-2431                     & Y & N & L & Y & Y & N & N\\
  24 & SSTc2d J162506.9-2350                     & Y & Y & \ldots  & Y & Y & N & Y\\
 26 & SSTc2d J182829.1+0027                    & Y & \ldots & H & Y & Y & N & T\\
  29 & SSTc2d J182911.5+0020                    & Y & Y & \ldots & Y & Y & N & N\\
  32 & SSTc2d J182935.6+0035                   & Y & Y & \ldots & Y & Y & N & Y\\
  33 & SSTc2d J182936.2+0042                   & Y & \ldots & L & Y & Y & N & T\\
  34 & SSTc2d J182944.1+0033                   & Y & Y & L & Y & N & N & N\\
  
\hline
    1 & SSTc2d J032835.0+3020$^\dagger$  & N & Y & L & Y & Y & N & N\\
    3 & SSTc2d J032903.9+3056$^\dagger$  & N & Y & \ldots & N & Y & Y & Y\\
  4 & SSTc2d J032924.1+3119$^\dagger$  & N & Y & L & N & Y & N & N\\
  6 & SSTc2d J033027.1+3028$^\dagger$  & N & Y & L & Y & Y & N & Y\\
  7 & SSTc2d J033035.5+3115$^\dagger$  & N & Y & L & N & Y & N & Y\\
  8 & SSTc2d J033038.2+3032$^\dagger$  & N & Y & \ldots & N & Y & N & Y\\
  9 & LZK 21                                                        & N & Y & L & Y & Y & N & Y\\
  13 & Cl* IC 348  LRL 265         & N & \ldots & L & Y & Y & N & T\\
  14 & Cl* IC 348  LRL 110                              & N & Y & L & Y & N & Y & N\\
  15 & LkH$\alpha$ 329                                   & N & Y & L  & Y & Y & N & Y\\
  16 & Hn 24                                                       & N & N & \ldots & Y & Y & N & Y\\
  20 & SSTc2d J162148.5-2340                     & N & Y & L & Y & Y & N & N\\
  21 & SSTc2d J162221.1-2304$^\dagger$& N & Y & L  & Y & Y & N & N\\
  23 & SSTc2d J162332.8-2258                     & N & N & \ldots  & Y & N & N & Y\\
  25 & DoAr 21                                                   & N & N & L & N & N & Y & N\\
  27 & SSTc2d J182858.1+0017                    & N & N & \ldots & N & N & Y & N\\
  28 & SSTc2d J182907.0+0038                    & N & \ldots & \ldots & Y & N & Y & N\\
  30 & SSTc2d J182915.6+0039                    & N & Y & \ldots & Y & Y & N & Y\\
  31 & SSTc2d J182915.6+0039$^\dagger$& N & \ldots & \ldots & Y & Y & N & N\\
  35 & SSTc2d J182947.3+0032                   & N & Y & L & Y & Y & N & N\\
\hline
\end{tabular}
\tablecomments{For clarity, cold disks and other objects are shown above and below the horizontal line, respectively. Y indicates feature is present, N indicates that the feature is not present, and T indicates a tentative detection.}
\tablecomments{$^1$: The classification of Sz 84 (\# 17) as a cold disk or an extended source is ambiguous but we keep it in assuming
it might be both.}
\tablecomments{$^\dagger$: Close to edge-on according to SED fit and luminosity below Main Sequence. In these cases, we
 do not classify as cold disks, since the hole size determination from the SED fit is highly uncertain.}

}
\end{table}

\clearpage

\vspace{-1cm}

\vspace{-1cm}

\vspace{-1cm}

\vspace{-1cm}

\clearpage

\clearpage

\clearpage

\end{document}